\PassOptionsToPackage{pdftex,dvipsnames}{xcolor}  
\documentclass[sigconf]{acmart}

\acmConference[SIGKDD'18]{Special Interest Group on Knowledge Discovery and Data Mining}{August 18}{London, England}

\setcopyright{rightsretained}

\usepackage{amsmath}
\usepackage{amsfonts}
\usepackage{amssymb}
\usepackage{amsthm}
\usepackage{balance}

\let\emptyset\varnothing

\usepackage[ruled,vlined]{algorithm2e}

\DontPrintSemicolon
\SetKwComment{tcp}{$\triangleright$ }{}
\SetVlineSkip{0cm}
\SetKwInOut{Param}{Param}
\SetKwInOut{Input}{Input}

\usepackage{graphicx}

\usepackage{subfigure}
\usepackage{wrapfig}

\usepackage{colortbl}
\usepackage{multicol}
\usepackage{multirow}
\renewcommand{\mid}{:}

\usepackage[font=small]{caption}
\usepackage{xargs}

\usepackage[colorinlistoftodos,prependcaption,textsize=tiny]{todonotes}
\newcommandx{\unsure}[2][1=]{\todo[linecolor=red,backgroundcolor=red!25,bordercolor=red,#1]{#2}}
\newcommandx{\change}[2][1=]{\todo[linecolor=blue,backgroundcolor=blue!25,bordercolor=blue,#1]{#2}}
\newcommandx{\info}[2][1=]{\todo[linecolor=OliveGreen,backgroundcolor=OliveGreen!25,bordercolor=OliveGreen,#1]{#2}}
\newcommandx{\improve}[2][1=]{\todo[linecolor=Plum,backgroundcolor=Plum!25,bordercolor=Plum,#1]{#2}}
\newcommandx{\inline}[2][1=]{\todo[inline,#1]{#2}}

\usepackage{pifont}
\newcommand{\cmark}{\text{\ding{51}}}
\newcommand{\xmark}{\text{\ding{55}}}

\definecolor{darkgreen}{rgb}{0,0.4,0}
\usepackage{hyperref}

\newcommand{\sysname}{{\sc gsaNA}\xspace}

\usepackage{tikz}
\usetikzlibrary{decorations.pathreplacing,calc}
\newcommand{\tikzmark}[1]{\tikz[overlay,remember picture] \node (#1) {};}

\newcommand*{\AddNote}[4]{%
    \begin{tikzpicture}[overlay, remember picture]
        \draw [decoration={brace,amplitude=0.5em},decorate,ultra thick,black]
            ($(#3)!(#1.north)!($(#3)-(0,1)$)$) --
            ($(#3)!(#2.south)!($(#3)-(0,1)$)$)
                node [align=center, text width=2.5cm, pos=0.5, anchor=west, color=black] {#4};
    \end{tikzpicture}
}%

\DeclareMathOperator*{\argmin}{argmin}
\DeclareMathOperator*{\argmax}{argmax}

\begin{document}


\title{An Iterative Global Structure-Assisted Labeled Network Aligner}

\author{Abdurrahman Ya\c{s}ar}
\affiliation{%
  \institution{School of Computational Science and Engineering}
  \streetaddress{Georgia Institute of Technology}
  \city{Atlanta}
  \state{Georgia}
  \postcode{30332}
}
\email{ayasar@gatech.edu}

\author{\"Umit V. \c{C}ataly\"urek}
\affiliation{%
  \institution{School of Computational Science and Engineering}
  \streetaddress{Georgia Institute of Technology}
  \city{Atlanta}
  \state{Georgia}
  \postcode{30332}
}
\email{umit@gatech.edu}

\begin{abstract}
Integrating data from heterogeneous sources is often modeled as merging graphs.
Given two or more
 ``compatible'', but not-isomorphic graphs, the first step
is to identify a {\em graph alignment}, where a potentially partial mapping
of vertices between two graphs is computed. A significant portion of the
literature on this problem only takes the global structure of the input graphs
into
account. Only more recent ones additionally use vertex and edge attributes to
achieve a more accurate alignment. However, these methods are not designed to
scale to map large graphs arising in many modern applications.
We propose a new iterative graph aligner, \sysname,
that uses the global structure of the graphs to significantly reduce the problem
size and align large graphs with a minimal loss
of information. Concretely, we show that our proposed technique is highly
flexible, can be used to achieve higher recall, and it is orders of magnitudes
faster than the current state of the art techniques.
\end{abstract}

\maketitle

\section{Introduction}
\label{sec:introduction}

The past decade witnessed unprecedented growth in the collection of data on
human activities, thanks to a confluence of factors including relentless
automation, exponentially reduced storage costs, electronic commerce,
geo-tagged personal technology devices and social media. This provides an
opportunity and a challenge to integrate heterogeneous sources of data and
collectively mine it. Many of these datasets are semi-structured or
unstructured, and naturally, can be best modeled as graphs. Hence, the problem
can be stated as integrating, or {\em merging} graphs coming from multiple
sources. The focus of this paper is merging two graphs.

Merging two graphs involves identifying each vertex in a graph with a
corresponding vertex (i.e., representing the same entity) in the other graph,
whenever such corresponding vertices exist. This problem, known as {\em graph
alignment}, is a well-studied problem that arises in many application areas
including computational biology~\cite{uetz2000nature,ito2001pnas},
databases~\cite{melnik2002icde},
computer vision~\cite{viola1997ijcv},
and network security and privacy~\cite{syed2011ftcit}. This is a challenging
problem as the underlying sub-graph isomorphism problem known to be NP-Complete~\cite{klau2009bmc}.
Once an alignment is identified, the final graph merging is a linear time
operation, hence we focus the subsequent discussion on the alignment
problem.

In the context of biological networks, such as protein protein interaction
(PPI) networks, graphs are smaller in size and usually, there is a high
structural similarity. We will show that the performance of the methods
for aligning such networks, both in terms of the solution quality and execution
time, needs significant improvement to handle the graphs that are of interest
in this work, which are much larger and highly irregular.
Some recent studies~\cite{zhang2016kdd} align more complex graphs,
where vertices and edges are associated with other metadata, such as types
and attributes.
The largest number of vertices tested in these
studies is only in the order of tens of thousands, as the algorithms are of high time complexity. In
addition, many of these studies rely on a sparse similarity
matrix~\cite{bayati2009icdm,klau2009bmc,zhang2016kdd,koutra2013icdm}
whose computation
requires quadratic (in terms of the
number of vertices) run time.
Koutra et al.~\cite{koutra2013icdm} try to overcome this problem by grouping the
vertices of the two graphs using their degrees. However, if the graphs are not
isomorphic or pseudo-isomorphic, this kind of an approach leads to large errors.

Our primary goal is to develop a scalable algorithm to
align two large graphs. The graphs are assumed to be ``similar'' but not
isomorphic, in other words,  they have different number of vertices and
edges, and adjacency structure of corresponding vertices might be different.
Graphs have additional metadata, such as types and attributes, on the
vertices and edges.

We propose a novel, fast network alignment algorithm, \sysname,
for the graph alignment problem. We take a divide-and-conquer approach and
partition the vertices into buckets. We then compare
the vertices of the first graph in a bucket with the vertices of the second graph that
are in the same bucket. The novelty of the proposed approach is to use the
global structure of the graph to partition the
vertices into buckets. The intuition behind \sysname is that for two vertices
$u$ and $v$ to map each other, they should be positioned in a ``similar
location'' in both graphs. To define the notion of ``similar location'', we
identify some {\em anchor vertices} in the graphs. These are reference
vertices that are either known to be true mappings or most likely to be.
We use each vertex's distance to a set of predetermined
anchors as a feature. We further use these distances to partition the problem
space, to reduce the computational complexity of the problem.



The contributions of this paper are as follows:

\begin{itemize}
\item We propose a global structure-based vertex positioning
technique to partition the vertices into buckets, which reduces
the search space of the problem.

\item We present an iterative algorithm to solve the
graph alignment problem by incrementally mapping vertices of input graphs.
At each step of the algorithm, similarity scores between vertices can take
the advantage of newly discovered high-quality mappings.

\item We propose generic similarity metrics for computing the
similarity of the vertices using structural properties and any
additional metadata available for vertices and edges.
\end{itemize}

Our experimental results show that our proposed algorithm, \sysname,
produces about $1.4\times$ better recall than {\tt Final}~\cite{zhang2016kdd},
$5\times$ better recall than {\tt IsoRank}~\cite{singh2007pairwise}, and
$2.6\times$ better recall than {\tt NetAlign}~\cite{bayati2009icdm},
$2.7\times$ better recall than {\tt Klau}~\cite{klau2009bmc}, when
we don't consider pathological cases for {\tt NetAlign} and {\tt Klau}.
\sysname outperforms these algorithms a couple of orders of magnitudes
in the execution time.

\section{Graph Alignment and Merge}
\label{sec:problemdef}

A graph $G=(V,E,T_V,T_E,A_V,A_E)$ consists of a set of vertices $V$, a set of
edges $E$, two sets for vertex and edge types $T_V$, $T_E$, and two sets for
vertex and edge attributes $A_V$, $A_E$, where type and attribute sets can be
$\emptyset$.
An edge $e$ is referred as $e = (u, v) \in E$, where $u, v \in V$. The
neighbor list
of a vertex $u \in V$ is defined as $N[u] = \{v\in V\mid
(u, v)\in E\}$. When discussing two graphs,
we will use subscripts
$1$ and $2$ to differentiate them if needed, and ignore those subscript when the intent
is clear in the context. For example, $N_1[u]$ and $N_2[u']$ will
represent the neighbor lists of vertices $u$ and $u'$ in $G_1$ and $G_2$,
respectively.
Given a vertex $x \in V$ or an edge $x \in E$, $a[x]$
represents the set of attributes of $x$, and $t[x]$ represents the type of
$x$. In addition, $L_V$ and $L_E$ represent the list of existing vertex and
edge types respectively. We also use $\delta(u,v)$ to denote the breath-first
search (BFS) distance between vertices $u$, and $v$. $S$ represents the map
of a small number of pre-known anchor vertices between these two graphs.

Given two different graphs $G_1$ and $G_2$, the similarity score between two
vertices $u \in V_1$ and $v \in V_2$ is denoted
by $\sigma : V^2 \rightarrow \mathbb{R}$. We will discuss different
variations and components of $\sigma$ in more detail in Section~
\ref{sec:similarity}.
We define $\mu[u] : V_1 \rightarrow V_2 $ as an injective mapping, where $\mu
[u]=v$ represents mapping of $u \in V_1$ to $v \in V_2$. If there is
no map, also refereed as {\bf nil} mapping, we use $\mu[u] =
\bot$.
Table~\ref{tab:TableOfNotation} displays the notations used in this paper.

\begin{table}[tbp]
\centering 
\begin{tabular}{r  l}

\textbf{Symbol} & \textbf{Description}                \\
\hline
$V$             & Vertex set \\
$E$             & Edge set \\
$T_{V}$         & Vertex type set \\
$T_{E}$         & Edge type set \\
$t[x]$          & Type of the vertex $x \in V$ or the edge $x\in E$\\
$A_{V}$         & Vertex attribute set  \\
$A_{E}$         & Edge attribute set \\
$a[x]$          & Attribute of the vertex $x \in V$ or the edge $x \in E$\\
$L_{V}$         & List of vertex types \\
$L_{E}$         & List of edge types \\
$N_i[u]$        & Neighbor list of vertex $u$ in graph $G_i$\\
$\delta(u,v)$   & Distance between $u, v\in V$ \\
$\sigma(u,v)$   & Similarity score for $u\in V_1$ and $v \in V_2$ \\
$\mu[u]$        & Mapping of $u \in V_1$ in $V_2$ \\
$S = S_1 \cup S_2 $  & Anchor (seed) set where $S_2 = \{v \mid
\mu[u]=v, v\in S_1 \}$ \\
\hline
\end{tabular}
\caption{Notations used in this paper.}
\label{tab:TableOfNotation}
\end{table}

\begin{definition}[Graph Alignment Problem]
Given two graphs $G_1=(V_1, E_1, \dots )$ and $G_2=(V_2, E_2, \dots )$,
the graph alignment problem is to find an
injective mapping that maximizes:
\begin{equation}
\sum_{\forall u, \mu[v]\neq \bot} \sigma(u, \mu[u]).
\end{equation}
\end{definition}

As we will discuss in Section~\ref{sec:similarity}, $\sigma$ can also
recursively depend on $\mu[v]$, hence optimization problem in hand is more
complex than the standard maximum weighted graph matching~\cite{west2001prentice}.

\begin{figure}
\centering
\includegraphics[width=0.90\linewidth]{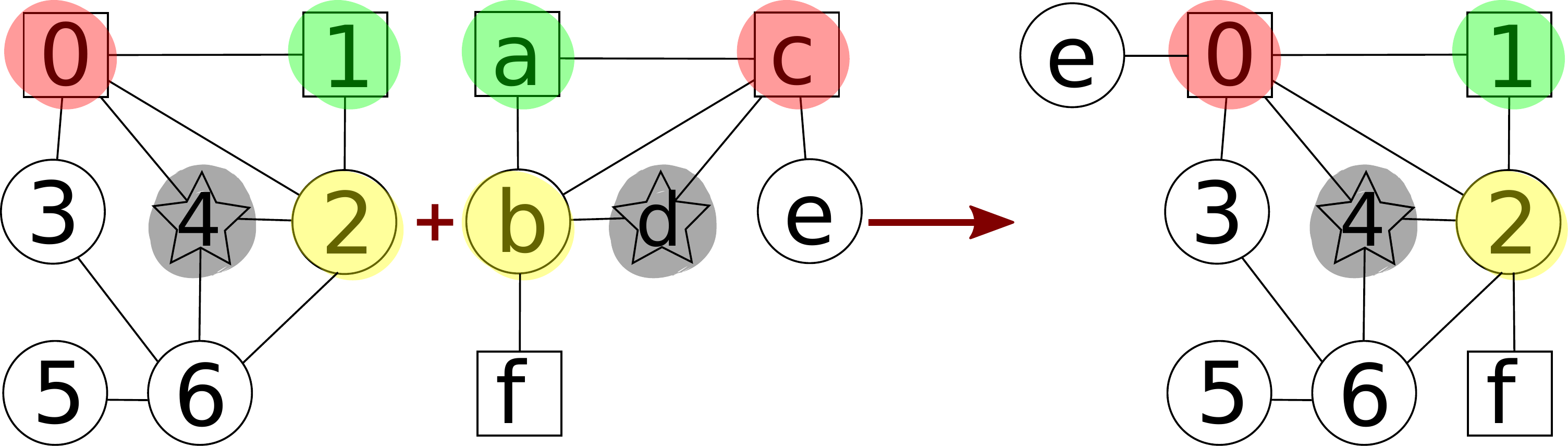}
\caption{A toy example showing graph merge problem.}
\label{fig:modelB}
\end{figure}

Given two graphs $G_1$ and $G_2$ (see Figure~
\ref{fig:modelB} for a toy
example), there may be
some vertices which are different and should not mapped.
For instance, consider the problem of merging two social networks, such as
Facebook and Twitter, although they have different purposes and
different structures, an important portion of their users have an account in
both networks, while the others do not.

\section{Iterative Global Structure Assisted Network Alignment}
\label{sec:gsana}

\begin{figure}[ht]
\centering
\includegraphics[width=\linewidth]{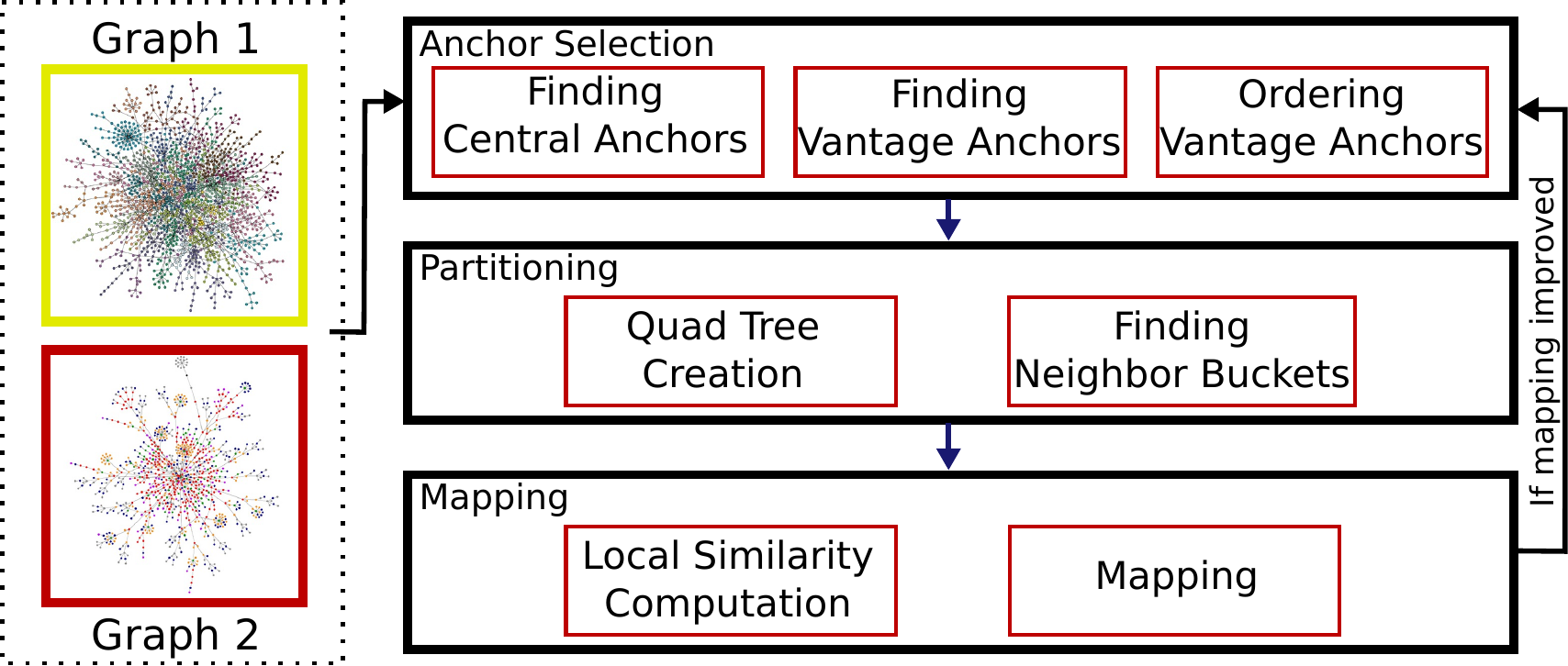}
\caption{Overview of the \sysname algorithm.}
\label{fig:overview}
\end{figure}

Figure~\ref{fig:overview} presents an overview of the proposed \sysname
algorithm. As illustrated in the figure, \sysname is composed of three
phases that are executed iteratively until a stable solution is found. Below we
give a high-level overview of each of these phases, then in the following
subsections, we discuss them in detail.

\begin{figure}[ht]
\centering
\subfigure[]{\includegraphics[width=0.55\linewidth]
{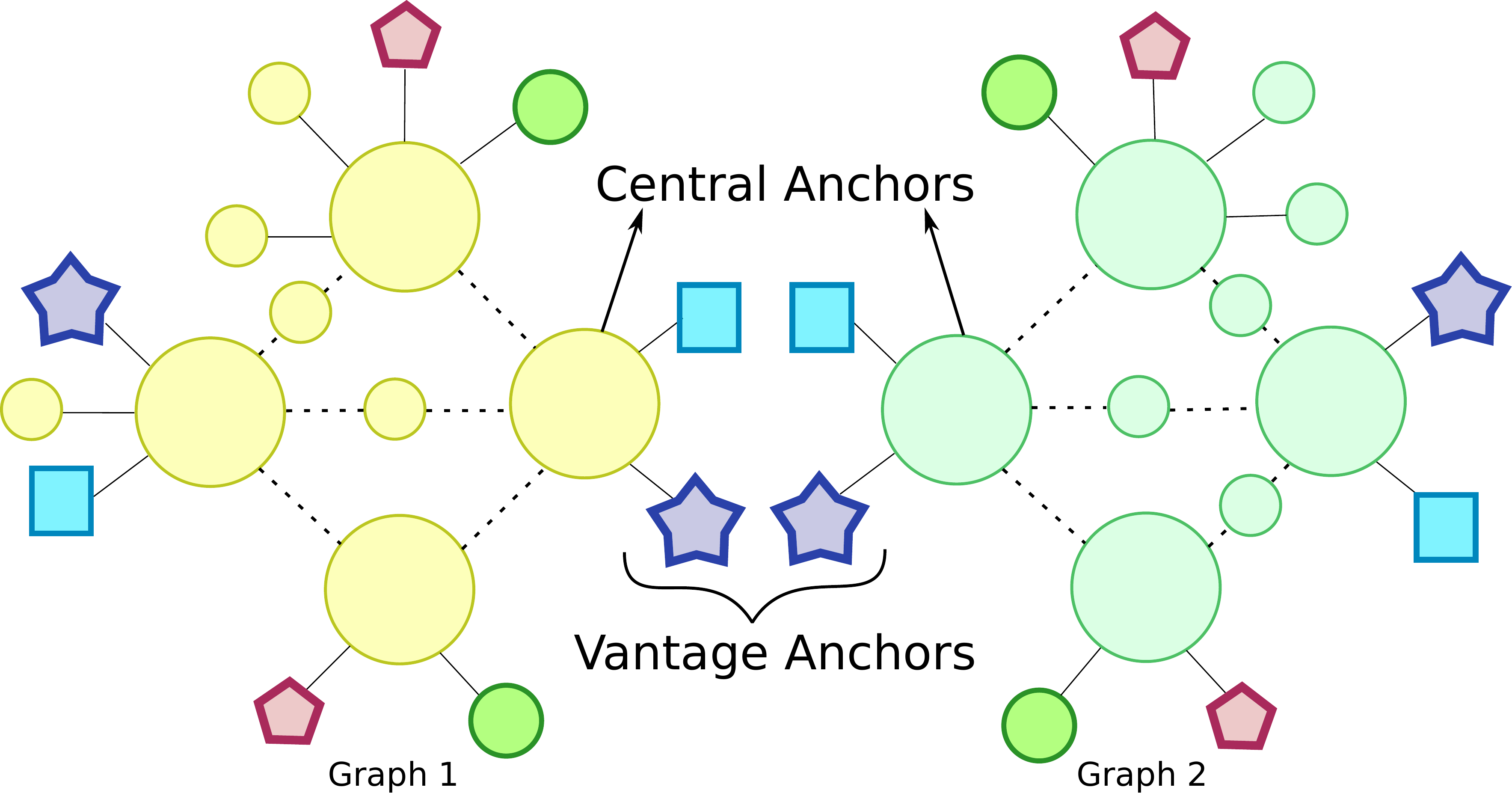}\label{fig:va}}
\subfigure[]{\includegraphics[width=0.42\linewidth]{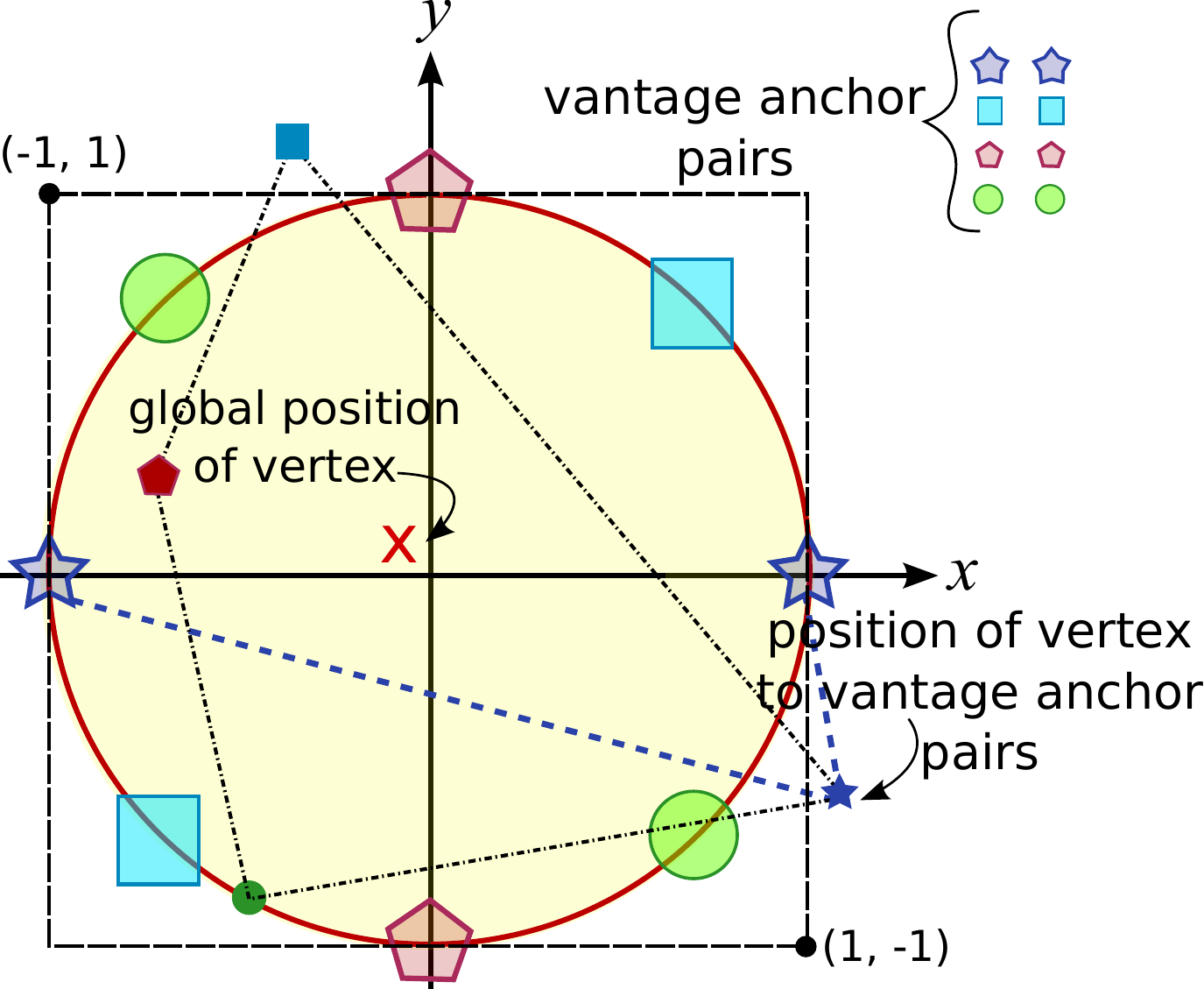}
\label{fig:gvm}}
\caption{(a) Vantage anchor selection, (b) positioning of vantage anchors to
unit circle, and vertex position computation.}
\label{fig:va-gvm}
\end{figure}

\paragraph{Anchor Selection:}
Anchors are a small subset of vertices whose mappings are known. These
anchors can be given
by the user or they can be computed by \sysname at the beginning.
Our goal here is to identify a smaller subset of anchors, that we
call {\em vantage anchors}, that can be used as reference points in the
rest of the algorithm. This is done in three steps. First, for a given set of
input anchors, \sysname computes the {\em central anchors} in both graphs.
Second, the remaining anchors are assigned to the closest central anchor.
This helps us to classify anchor vertices. If two anchors are close to
the same central anchor, then they cannot be good candidates to distinguish
the vertices.
\sysname chooses vantage anchors from these assigned anchors. Finally,
\sysname pairs each vantage anchor with the most
``distant ones'', orders and places them onto a unit circle to be used in the next phase.
Figure~\ref{fig:va-gvm} illustrates this process.

\paragraph{Partitioning:}
In this phase, vertices are partitioned into {\em buckets} on a $2D$ plane
using their distances to the vantage anchors pairs. The intuition is
that for a vertex $u$ to be mapped to a vertex $v$ in the other graph, their
distances to the selected vantage anchor pairs should be similar. Hence, in
this step, first, each vertex's distances to vantage anchors are computed.
Then, for each pair of vantage point position of the vertex on this $2D$
plane is computed. These positions define a polygon for each vertex. Finally,
a single location is calculated by computing the centroid of this polygon.
These final locations are used to partition the vertices into buckets. Due to
the skewed and irregular structure of the graphs we expect the
distribution of the positions will be skewed on this $2D$ plane. Therefore
\sysname partitions the plane with {\em quad-trees}~\cite{finkel1974quad}.

\paragraph{Mapping:}
The last phase of the algorithm is to compute pairwise similarity among the
vertices of the two graphs that fell into the same bucket. Then compute a
potentially partial mapping. The process is repeated for all non-empty buckets.

\paragraph{Iterations}

The recall of \sysname depends on the quality of the selected
vantage anchors pairs. After computing a mapping, we have more information
available for the alignment. By leveraging this information, we can
recompute the vantage anchors, partition the vertices and map them.
This way, we can choose better vantage anchors and decrease the number of
false hits. We iteratively do these steps until the mapping is stable, i.e.,
it does not change
more than a small fraction (we used 2\% in our experiments). As initial anchors,
we pick the highest scored
mappings, and we double the number of anchors we use in each iteration, but
we put an upper bound on that (1,000), and go back to initial anchors if we
exceed that. Alg.~
\ref{alg:gsaNA} presents
high-level pseudo-code of \sysname.


\begin{algorithm}[ht]
\begin{small}
  $\mu \leftarrow \mu_a$, $\mu_p \leftarrow \emptyset$  \;
  $a \leftarrow |S_1|$\tcp*[r]{Initialize $a$ as the size of the anchors
  set}
  $k \leftarrow 3$\tcp*[r]{Initialize $k$ number of top similar vertices per vertex}
  $n \leftarrow 20$\tcp*[r]{Initialize maximum number of iterations}
  $\epsilon \leftarrow 1.02$\tcp*[r]{Minimum changes for next iteration}

  \While{$(n>0)$ and $(|\mu| \; / \; |\mu_p| > \epsilon)$ }{
    $\mu_p \leftarrow \mu$\tcp*[r]{Store current mapping in $\mu_p$}
    \tcp*[l]{Computation of shortest paths from seed anchors}
      \For{{\bf each} $u \in S_1$}{
      \tcp*[l]{For each ``new'' seed anchor perform a BFS}
        \If{ $\delta(u, .)$ is not computed before } {
          $\delta(u, .) \leftarrow BFS(G_1, u)$,
          $\delta(\mu[u], .) \leftarrow BFS(G_2, \mu[u])$\;
        }
      }

    $S_C \leftarrow$ {\sc findCentralAnchors}$(G_1, S_1, \delta, 1)$
    \tikzmark{topAS}\;
    $S_V \leftarrow$ {\sc findVantageAnchors}$(S_1 \setminus S_C, S_C, \delta)$
    \tikzmark{right} \;
    $O_V \leftarrow ${\sc pairAndOrder}$(S_V, \delta)$ \tikzmark{bottomAS}\;

    $Q \leftarrow$ {\sc QTree}$((-1,1), (1,-1))$ \tikzmark{topVP}\;
    $T \leftarrow$ {\sc insertVertices}$( V_1 \cup V_2, O_V, Q, \delta )$
    \tikzmark{bottomVP}\;

    $P \leftarrow$ {\sc topSimilars}$(T, k, \sigma)$ \tikzmark{topM}\;
    $\mu \leftarrow$ {\sc map}$(P, \mu, \sigma)$ \tikzmark{bottomM}\;

    \tcp*[l]{Append highest similar vertices as new new anchors}

    \If{$a>1000$} {
      $S_1 = \{ u \mid \mu_a[u] = v \}$\;
      $a \leftarrow |S_1|$ \;
    }

    \For{$i=1$ to $a$} {
      $S_1 \cup \{u\}$, where $u \leftarrow \argmax_{u \in V_1 \setminus S_1}{\sigma(u, v)}$\;
    }
    $a \leftarrow 2\times a$ \;
    $n \leftarrow n-1$\;
  }

  \Return $\mu$
  \AddNote{topAS}{bottomAS}{right}{ \textbf{Anchor Sel.} }
  \AddNote{topVP}{bottomVP}{right}{ \textbf{Partitioning} }
  \AddNote{topM}{bottomM}{right}{ \textbf{Mapping} }
  \caption{{\sc gsaNA($G_{1}, G_{2}, S_{1}$)}}\label{alg:gsaNA}
\end{small}
\end{algorithm}

\subsection{Anchor Selection}
\label{subsec:anchorselection}

The size of the anchor set plays important role in \sysname. We cannot
request a complete mapping, but we
need a few good anchors to start with.
If an anchor set is not given by the user,
we set $|S| = 4 \times \log(\max(|V_1|,|V_2|))$,
and bootstrap the algorithm by
finding $2 \times |S|$ highest degree vertices in both graphs, and then by computing
an initial mapping based on the similarity
scores among them (see Section~\ref{sec:similarity}).

Given a centrality metric, we define set of {\em central anchors}
$S_C$ as the $l = \log(|S|)$ vertices within the anchor set $S$ which
have the highest centrality measures, and not ``too close to each other''.
Among many centrality measures~\cite{linton1978sn}, we use the degree
centrality.
Alg.~\ref{alg:cafinding} presents the pseudo-code for this step.

\begin{algorithm}[ht]
\begin{small}
  \tcp*[l]{$\delta$ is a distance function, $t$ is distance threshold}

  \tcp*[l]{First find non-close anchors}
  $S' \leftarrow \emptyset $ \;
  \For{{\bf each} $u \in S$}{
      \If{$\forall v\in S', \delta(u,v) > t$}{
          $S' \leftarrow \{u\}$ \;
      }
  }

  $l \leftarrow \log(|S|)$\tcp*[r]{Size limit of central anchors}
  \tcp*[l]{Get $l$ anchors with the highest degree}
  $C \leftarrow \emptyset$ \;
  \For{$i=1$ to $l$}{
      $C \leftarrow \{u\}$, where $u \leftarrow \argmax_{u\in S'\setminus C} |N
      [u]|$ \;
  }
  \Return $C$
  \caption{{\sc findCentralAnchors($G, S, \delta(\cdot,\cdot), t$)}}
  \label{alg:cafinding}
\end{small}
\end{algorithm}

\sysname uses the central anchors to classify the rest of the anchors, where a
subset of them is selected as {\em vantage anchors}.
\sysname uses the vantage anchors as the main reference
points to partition the vertices of the graphs. Pseudo-code for finding vantage
anchors is presented in Alg.~\ref{alg:assign}. To identify them, for each
non-central anchor, we first find the closest central anchor and assign
non-central anchor to it.
Then in order to evenly distribute vantage anchors over the graph, we limit
the number of vantage anchors per central anchor, with the minimum number of
assigned anchors to any central anchor. After, for each central anchor, among the
assigned non-central anchors, we select the anchors that are farthest to it.
Here, when needed, we break the ties by picking the
anchor that is farthest than all other central anchors (this is not displayed
in the algorithm).

\begin{algorithm}[ht]
\begin{small}
  $\forall c\in C, S_V'[c] \leftarrow \emptyset $\tcp*[r]{Initialize vantage
  anchor list}

  \For{{\bf each} $u \in S'$}{
    $S_V'[c] \leftarrow  \{u\}$, where $c \leftarrow \argmin_{c \in C} \delta
(u, c)$\;
  }

  $ a \leftarrow \min_{c\in C} |S_V'[c]|$ \tcp*[r]{limit num. of vantage anchors}

  $S_V \leftarrow \emptyset$ \;
  \For{{\bf each} $c \in C$} {
    \For{$i=1$ to $a$}{
      $S_V \leftarrow \{u\}$, where $u \leftarrow \argmax_{u \in S_V'[c]} \delta
(u, c)$ \; 
    }
  }

  \Return $S_V$
  \caption{{\sc findVantageAnchors($S', C, \delta$)}}
  \label{alg:assign}
\end{small}
\end{algorithm}

Once anchors are assigned to the central anchors, we call Alg.~\ref{alg:order} to pair
the assigned anchors using the distance function $\delta$ and order them. Each vantage anchor is paired with the farthest
vantage anchor. Then, one of the pairs is selected as the first pair. The rest
of the pairs are ordered based on the distance of their first vertex to
previous pair's first vertex.

\begin{algorithm}[ht]
\begin{small}
  $cnt \leftarrow 0$ \tcp*[r]{Initialize counter for ordered pairs}
  \For{{\bf each} $u \in S_V$} {

    $v \leftarrow \argmax_{v\neq u \in S_V} \delta(u, v)$ \;
    $O_V[cnt] \leftarrow [u, v]$ \;
    $cnt \leftarrow cnt + 1$ \;
    $S_V \leftarrow S_V \setminus \{u, v\}$ \;
    \If{$|S_V| \leq 1$}{
      {\bf break} \;
    }
  }

  \For{$i=2$ to $cnt$} {
    $j \leftarrow \argmin_{i \leq j \leq cnt} \delta(O_V[i-1][1], O_V
    [j][1])$\;
    {\bf swap}($O_V[i], O_V[j])$
  }

  \Return $O_V$
  \caption{{\sc pairAndOrder($S_V, \delta$)}}
  \label{alg:order}
\end{small}
\end{algorithm}

\subsection{Partitioning}
\label{sec:partitioning}

The ordering of the vantage anchors are used to place them on a unit circle. The
first pair is assumed to be ``placed'' at (1, 0) and (-1, 0), then second pair
is placed on the unit circle with rotating $\pi / (|S|/2)$ in counter-clockwise
(see Figure~\ref{fig:gvm}).
Then, we compute the ``position'' of a vertex by placing the vertex
as the corner of a right-angle triangle that is composed of the vertex and the
vantage anchor pairs. So its distance to vantage anchors is scaled with the
distance between vantage points and a point is computed using simple
trigonometric functions. We repeat this process for every vantage anchor pairs. We
then compute a final global position for the vertex as the centroid of the
polygon defined by these locations as corners. The
algorithm for this computation is displayed in Alg.~\ref{alg:globalpos}.
To partition vertices of the two input graphs,  we simply compute a
global position of each vertex using this algorithm, and then insert them
into a quadtree.
If a bucket exceeds
pre-defined size limit, $B$, then that bucket is split into four. This
continues until all of the vertices are inserted.

\begin{algorithm}[ht]
\begin{small}

  $\theta_i \leftarrow \theta \leftarrow \pi /|O_V|$\;
  $poly \leftarrow \emptyset$ \;
  \For{$p  \in O_V$}{
    \tcp*[l]{distances between anchors and $u$}
    $a \leftarrow \delta(u, p[0])$, \ \
    $b \leftarrow \delta(u, p[1])$, \ \
    $c \leftarrow \delta(p[0], p[1])$ \;
    \tcp*[l]{compute angle between $p[0]$ and $u$}
    $\alpha \leftarrow arccos(a^2+c^2 - b^2) / (2 \times ac)$ \;
    \tcp*[l]{compute $x$ and $y$ coordinates of $u$}
    $x \leftarrow p[0].x - a \times cos(\alpha)$, \ \ \
    $y \leftarrow a \times sin(\alpha)$ \;

    $poly.insert( Rotate((x, y), \theta_i) )$ \tcp*[r]{rotate and insert into
    $l$}
    $\theta_i \leftarrow rot+\theta$\;
  }

  \Return $Centroid(poly)$
  \caption{{\sc getVertexPosition($u, O_V, \delta $)}}
  \label{alg:globalpos}
\end{small}
\end{algorithm}

\subsection{Mapping}
\label{sec:mapping}

Mapping is the third phase of our iterative algorithm, and the goal
is to compute a mapping between vertices of the two graphs. As we
have stated in Sec.~\ref{sec:problemdef}, mapping step can be modeled as a
maximum weighted bipartite graph matching~\cite{west2001prentice} problem.
However, since our overall algorithm is an iterative one, we will be sensitive
and only finalize mappings that are most likely be the true mappings in each
iteration, hoping that in further iterations, with more mapping information
becomes available, similarity scores will reflect those and we can make
better mappings.
Our mapping algorithm (Alg.~\ref{alg:compsim}) starts with computing similarity
scores for
each vertex $v \in V_2$ in a non-empty bucket $B$ of the quadtree with vertex
$u \in V_1$ that is either in the same bucket, or in one of the ``neighboring''
buckets. We check neighbor buckets to make sure that vertices are close to the
border of the buckets are handled appropriately. After we identified top $k$
similar vertices (stored in $P[v]$), we compute a mapping, potentially a partial
one, using Alg.~\ref{alg:map}. For each vertex, $u\in V_2$, we get $v\in V_1$
such that $v$ has the highest similarity score among the other candidates for $u$.
Then we check the previously assigned mapping of $v$. If it had no mapping
or previous mapping similarity score was less than what we have now, we
mark $v$ as mapped to $u$. We repeat this procedure until there is no change
in mapping. In short, our greedy mapping algorithm only considers the top $k$
best mappings, and only accept mapping both vertices agrees that their best
``suitor'' is each other.

\begin{algorithm}[ht]
\begin{small}
  \tcp*[l]{For each vertex keep a priority list with top $k$ elements.}
  $P[v] \leftarrow \emptyset$, for $\forall v\in V_2$\;

  \For{{\bf each} non-empty $B \in Q$}{
    \For{{\bf each} $v \in B \wedge v \in V_2$} {
      \For{{\bf each} $u \in Neig(B) \wedge u \in V_1$} {
        $P[v].insert(u)$  \tcp*[r]{Only keeps top $k$}
      }
    }
  }

  \Return $P$
  \caption{{\sc topSimilars($Q, k, \sigma$)}}
  \label{alg:compsim}
\end{small}
\end{algorithm}
\begin{algorithm}[ht]
\begin{small}

  $\mu_p \leftarrow \emptyset$ \;
  \While{$\mu_p \neq \mu $}{
    $\mu_p \leftarrow \mu$ \;
    \For{{\bf each} $u\in V_2$ where $P[u] \neq \emptyset$}{
      $v \leftarrow P[u].pop()$ \tcp*[r]{Pop the current best for $u$}
      \If{$\mu[v] = \bot$ {\bf or} $\sigma(v,u) > \sigma(v, \mu[v])$} {
        $\mu[v] = u$\;
      }
    }
  }
  \Return $\mu$
  \caption{{\sc map($P, \mu, \sigma$)}}
  \label{alg:map}
\end{small}
\end{algorithm}

\subsection{Similarity Metrics}
\label{sec:similarity}

Our similarity score is composed of multiple components, some only depend
on graph structure, some depends also on the additional metadata (types and
attributes).
Given two graphs $G_1$ and $G_2$ the similarity of two vertices, $u \in V_1$
and $v \in V_2$, our composite score is a simple average of six components  as
follows:

\begin{equation}
\sigma(u,v) = \frac{\tau}{6} \times ( \alpha + \Delta + \tau_V + \tau_E + C_V +
C_E )
\end{equation}

\noindent where Table~\ref{tab:metrics} lists the description of each
component in this equation. Using these metrics we try to cover
different graph characteristics which may help to increase final recall.
Graph structure scores will be always available, and we include
additional components when they are available.
For example, when there is no additional metadata is available, our similarity
score will reduce to average of two structural components:

\begin{equation}
\sigma(u,v) = \frac{1}{2} \times ( \alpha + \Delta ).
\end{equation}

\begin{table}[htbp]
\centering
\caption{Components of the similarity function.}
\label{tab:metrics}
\begin{tabular}{|r c l|} \hline
{\bf Symbol} & &{\bf Description} \\ \hline
$\tau$   &  :&Type similarity\\
$\alpha$ & : &Anchor similarity \\
$\Delta$ & : &Relative degree distance~\cite{koutra2013icdm}\\
$\tau_V$ & : &$\#Same/\#Total$ types of adjacent vertices\\
$\tau_E$ & : &$\#Same/\#Total$ types of adjacent edges\\
$C_V$    & : &Vertex attribute similarity\\
$C_E$    & : &Edge attribute similarity\\ \hline
\end{tabular}
\end{table}

In a typed graph, types of the vertices are very important; for instance,
one would not want to map a \texttt{human} in a graph with a
\texttt{shop} in another graph. For this reason, we define {\em type
similarity}, $\tau$, as a boolean metric (1 or 0), and it checks if
the types of the vertices are the same or not.

{\em Anchor similarity}, $\alpha(u,v)$, is defined as the ratio of the number
of common anchors among the neighbors of $u$ and $v$ to the total number of
anchors in them. Formally, it is defined as:

\begin{equation}
\alpha(u,v)=\frac {|\{w \mid w\in N_1[u] \wedge \mu[w]\in N_2[v] \wedge w\in
S_1 \}|}
                  {|\{w\mid w\in N_1[u] \cup N_2[v]\wedge w\in S\}|}
\end{equation}
We borrow the {\em relative degree distance} from~\cite{koutra2013icdm}:

\begin{equation}
\Delta(u,v) = \Big( 1+\frac{ 2\times| |N_{1}[u]| - |N_{2}[v]| |}
                         {| |N_{1}[u]| + |N_{2}[v]| |} \Big)^{-1}
\end{equation}

The next two components takes type distributions of the neighboring vertices
and edges into account. Let us define $tc[u,l]$ as the number of neighbors
of $u$ of the type $l$, i.e., $tc[u,l]=|\{u'\mid u' \in N_i[u]\} \wedge t
[u']=l\}|$. Then we define {\em neighborhood vertex type similarity}, $\tau_V
(u,v)$, follows:

\begin{equation}
\tau_V(u,v)= \frac{\sum\limits_{l \in L_V } \min (tc[u,l], tc
[v,l])} {\sum\limits_{l \in L_V } \max ( tc[u,l], tc[v,l] ) }
\end{equation}

{\em Neighborhood edge type similarity}, $\tau_E$, is also defined
in a similar way; we omit the equation for simplicity.

When the vertices and/or edges have attributes, we take those into account with
attribute similarity metrics. Formally we define {\em vertex attribute
similarity}, $C_V$, like a weighted, generalized Jaccard similarity,
such that:

\begin{equation}
C_V(u,v) = \frac{\sum_{c \in a[u] \cap a[v]} \min{w(c)}}
                {\sum_{c \in a[u] \cup a[v]} \max{w(c)}}
\label{eq:vattr}
\end{equation}
\noindent where $w(c)$ represents the weight of a non-numeric attribute.

When edge attributes are are non-numeric, we also define {\em edge attribute
similarity}, $C_E$, very similar to Eq.~\ref{eq:vattr}. When attributes are numeric,
we define $C_E$ as follows:

\begin{equation*}
C_E(u,v) = \frac{\sum_{u' \in N_{1}[u]} \sum_{v' \in N_{2}[v]} close((u, u'), (v,
v'))} {|N_{1}[u]| \times |N_{2}[v]|}
\end{equation*}
\noindent where $close(e, e')$ is a boolean function, and it is 1 when value
$|val
(e)-val(e) | < \varepsilon$ for all numeric attributes, and 0 otherwise. Here,
$val(e)$ denotes the value of edge attribute, and $\varepsilon$ is pre-determined threshold.

\section{Related Work}
\label{sec:related}

Solution methods proposed in the literature for graph alignment can be
roughly classified into four basic
categories~\cite{conte2004ijprai,elmsallati2016itcbb}: spectral
methods~\cite{singh2007pairwise,liao2009bioinf,patro2012bioinf,neyshabur2013bioinf},
graph structure similarity
methods~\cite{kuchaiev2010topological,milenkovic2010optimal,vesna2012graal,dogning2015binf,aladag2013bioinf},
tree search or tabu search
methods~\cite{chindelevitch2013optimizing,saraph2014bioinf,liu2007multiobjective,kpodjedo2014using},
and integer linear programming (ILP)
methods~\cite{klau2009bmc,kebir2011icprb,bayati2009icdm}.
All of these works have scalability issues.
Our algorithms
leverage global graph structure and reduces the problem space and augment that
with
semantic information to alleviate most of the scalability issues.

As an example of spectral methods, {\tt IsoRank}~\cite{singh2007pairwise}---one of
the earliest global alignment work
in computational biology---suggests an eigenvalue problem that approximates
the objective of finding the maximum common subgraph. After finding the vertex similarity matrix,
IsoRank finds the alignment by solving the maximum weighted bipartite matching.
IsoRank finds a 1/2-approximate matching using a greedy method, which aligns
pair of vertices in the order of highest estimated
similarity.
IsoRank was extended to multiple networks
in~\cite{liao2009bioinf}, where pairwise similarity matrices are computed and
an iterative spectral clustering is used to output set of vertices, one from each
network, that aligns with each other.
It is noted that it cannot handle more
than five networks.


In~\cite{klau2009bmc}, named as {\tt Klau} in our experiments, the problem of
finding the mapping with the maximum
score is posed as an integer quadratic program. It is solved by an integer
linear programming (ILP) formulation via a sequence of max-weight matching
problems. Authors use Lagrangian relaxation to solve this problem approximately
in a more reasonable time. However, the ILP based solutions will not scale to larger
problem sizes.

{\tt NetAlign}~\cite{bayati2009icdm}  formulates
the network alignment problem as an integer quadratic
programming problem to maximize the number of ``squares''. A near-optimal
solution is obtained by finding the maximum a posteriori assignment using belief
propagation heuristic and message-passing algorithms which yield near optimal
results in practice.
Another message passing network
alignment algorithm on top of belief propagation is proposed
by Bradde et al.~\cite{bradde2010epl}. In~\cite{koutra2013icdm} Koutra et al. propose to
align two bipartite graphs with a fast projected gradient descent algorithm which exploits
the structural properties of the graphs.

In a more recent work, Zhang et al. propose {\tt Final}~\cite{zhang2016kdd} to
solve attributed
network alignment problem. {\tt Final} extends the concept of {\tt IsoRank}~
\cite{singh2007pairwise}, and
make it capable to benefit from attribute information of the vertices and edges to
solve this problem. In addition to graph's vertex, edge and attribute sets
{\tt Final} adds an optional input called {\em prior knowledge matrix (H)}
in which each
entry gives likelihood to align two vertices.
{\tt Final} is one of the most recent works which solves
attributed graph alignment problem and
outperforms~\cite{singh2007pairwise,klau2009bmc,bayati2009icdm,koutra2013icdm}.



\begin{table*}[t]
\centering
\caption{Properties of the datasets. $\langle|N[x]|\rangle$ represents
average vertex
degree, and $|\mu|$ represent the size of ground truth mapping.}
\label{table:dataset}
\begin{tabular}{  l  
                  r  r  
                  r  r  r r 
                  r  
                  r  r  
                  r  
                  c  c  
                  }

\hline
\multicolumn{1}{l}{\textbf{Data Set}}  &
\multicolumn{1}{r}{$\boldsymbol{|V|}$}   &
\multicolumn{1}{r}{$\boldsymbol{|E|}$}   &
\multicolumn{1}{c}{$\boldsymbol{\langle|N[x]|\rangle}$}   &
\multicolumn{1}{c}{$\boldsymbol{\max(|N[x]|)}$}   &
\multicolumn{2}{r}{$\boldsymbol{|N[x]| < 3}$}   &
\multicolumn{1}{c}{$\boldsymbol{|\mu|}$} &
\multicolumn{1}{c}{ $\boldsymbol{|L_V|}$}&
\multicolumn{1}{c}{ $\boldsymbol{|L_E|}$} &
\multicolumn{1}{c}{ $\boldsymbol{|S_1|}$} &
\multicolumn{1}{c}{$\boldsymbol{A_V}$} &
\multicolumn{1}{c}{$\boldsymbol{A_E}$} \\ \hline
Douban-Online    & 3,906   & 16,328  & 4.18  & 124  &  1,467 & (38\%) & 
\multirow{2}{*}{1,118}  & 
\multirow{2}{*}{538}  & 
\multirow{2}{*}{2}  & 
\multirow{2}{*}{48}  & 
\multirow{2}{*}{\xmark} & 
\multirow{2}{*}{\xmark} \\ 
Douban-Offline    & 1,118   & 3,022   & 2.71  & 38  & 638 &(57\%)&    &   & \\ \hline
Facebook-1    & 4,038   & 88,234  & 21.86  & 696  &  173 &(4\%) &
\multirow{2}{*}{4,011}  &
\multirow{2}{*}{5}  &
\multirow{2}{*}{1}  &
\multirow{2}{*}{48}  &
\multirow{2}{*}{\xmark} &
\multirow{2}{*}{\xmark} \\
Facebook-2    & 4,438   & 79,411  & 17.89  & 615  &  196 &(4\%)&  &   & \\ \hline
Lastfm      & 15,436  & 32,638  & 2.11  & 1,952  &  13,961 & (91\%)&
\multirow{2}{*}{452}  &
\multirow{2}{*}{3} &
\multirow{2}{*}{3}  &
\multirow{2}{*}{56}  &
\multirow{2}{*}{\cmark} &
\multirow{2}{*}{\xmark} \\
Flickr      & 12,974  & 32,298  & 2.49  & 1,736  & 10,383 & (81\%)&   &   & \\
\hline
Myspace     & 10,733  & 21,767  & 2.03  & 326  &  10,120 &(94\%)&
\multirow{2}{*}{267}  &
\multirow{2}{*}{3}  &
\multirow{2}{*}{3}  &
\multirow{2}{*}{54}  &
\multirow{2}{*}{\cmark} &
\multirow{2}{*}{\xmark} \\
Flickr      & 6,714   & 14,666    & 2.18  & 1,278  &  5,836 &(87\%)&  &   & \\ \hline \hline
DBLP-17 (0)   & 59,006 & 665,800 & 11.28  & 2,322  &  3,098 &(5\%)&
\multirow{2}{*}{27,029} &
\multirow{2}{*}{1} &
\multirow{2}{*}{1}  &
\multirow{2}{*}{68}  &
\multirow{2}{*}{\cmark} &
\multirow{2}{*}{\cmark} \\
DBLP-14 (0)   & 43,936    & 368,983 & 8.40 & 1,782  &  3,248 &(7\%)&   &   & \\ \hline
DBLP-17 (1)   & 118,012 & 1,287,928 & 10.91  & 2,322  &  7,086 &(6\%)&
\multirow{2}{*}{60,902} &
\multirow{2}{*}{1} &
\multirow{2}{*}{1}  &
\multirow{2}{*}{68}  &
\multirow{2}{*}{\cmark} &
\multirow{2}{*}{\cmark} \\
DBLP-14 (1)   & 87,873    & 705,725 & 8.03 & 1,782  &  7,230 &(8\%)&   &   & \\ \hline
DBLP-17 (2)   & 236,025 & 2,232,274 & 9.46  & 2,322  &  17,364 & (7\%)&
\multirow{2}{*}{130,786}  &
\multirow{2}{*}{1} &
\multirow{2}{*}{1} &
\multirow{2}{*}{72} &
\multirow{2}{*}{\cmark} &
\multirow{2}{*}{\cmark} \\
DBLP-14 (2)   & 175,746 & 1,322,910 & 7.43  & 1,782 &  17,688 &(10\%)&  &   &\\ \hline
DBLP-17 (3)   & 491,719 & 4,089,071 & 8.31  & 2,322  &  51,035 &(10\%)&
\multirow{2}{*}{294,531}  &
\multirow{2}{*}{1} &
\multirow{2}{*}{1} &
\multirow{2}{*}{75} &
\multirow{2}{*}{\cmark} &
\multirow{2}{*}{\cmark} \\
DBLP-14 (3)   & 366,137 & 2,542,331 & 6.94  & 1,782  &  46,853 &(13\%)&   &   & \\ \hline
DBLP-17 (4)   & 983,438 & 6,685,519 & 6.80  & 2,322  &  148,408 &(15\%)&
\multirow{2}{*}{649,500}  &
\multirow{2}{*}{1} &
\multirow{2}{*}{1}   &
\multirow{2}{*}{79}  &
\multirow{2}{*}{\cmark} &
\multirow{2}{*}{\cmark} \\
DBLP-14 (4)   & 732,275 & 4,268,145 & 5.83  & 1,782  &  128,641 &(18\%)&   &   & \\ \hline
DBLP-17     & 1,966,877 & 9,059,634 & 4.61  & 2,322  & 616,386 &(31\%)&   &   & \\
DBLP-14     & 1,464,539 & 5,906,792 & 4.03  & 1,782  &  491,206 &(34\%)&
1,440,379  &
1 &
1 &
83 &
\cmark &
\cmark \\
DBLP-15     & 1,620,196 & 6,828,586 & 4.22  & 2,168  & 528,949 &(33\%)&
1,601,443  &
1 &
1 &
83 &
\cmark &
\cmark \\
DBLP-16     & 1,783,746 & 7,841,210 & 4.40  & 2,149  & 571,703 &(32\%)&
1,772,129  &
1 &
1 &
83 &
\cmark &
\cmark\\ \hline
\end{tabular}
\end{table*}

\section{Experimental Evaluation}
\label{sec:experimental}

In this section, we first present several experiments in order to identify the
performance trade-offs of the parameters of \sysname.

We then compare the performance of proposed \sysname algorithm
\sysname against four state-of-the-art
mapping algorithms: {\tt IsoRank}~\cite{singh2007pairwise},
{\tt Klau}~\cite{klau2009bmc}, {\tt NetAlign}~\cite{bayati2009icdm},
and {\tt Final}~\cite{zhang2016kdd}, each briefly described in the previous
section.
We also present performance of these algorithms and \sysname when there are
errors
in the graph structure or in the attributes.
In our experiments we used Matlab implementations of these algorithms
~\cite{zhang2016kdd-data,bayati2009icdm-repo}.

Experiments were carried out on machine that has 2 16-core Intel Xeon
E5-2683 2.10GHz processors, 512GB of memory, 1TB disk
space, running Ubuntu GNU/Linux with kernel 4.8.0. \sysname is
implemented in C++ and complied with GCC 5.4.

\subsection{Dataset}

We use real-world graphs obtained from~\cite{zhang2016kdd-data,ref:dblp-site,ref:dnd}.
We also generated different size of DBLP~\cite{ref:dblp-site}
graphs.
The properties of graphs are listed in Table~\ref{table:dataset} and we
briefly describe them below.

{\em Douban Online-Offline}~\cite{zhang2016kdd-data}:
These two graphs are extracted subnetworks of the original dataset~\cite{zhong2012kdd}. The original
dataset contains $50k$ users and $5M$ edges.
Both networks are constructed using users' co-occurrences
in social gatherings. In~\cite{zhang2016kdd} people are treated as, (i)~
'contacts' of each other if the
cardinality of their common event participations is between ten and twenty
times, (ii)~ 'friends' if the cardinality of their common event participation
is greater than 20.
The constructed offline and online network has 1,118 and 3,906 vertices
respectively.
The location of a user is used as the vertex attribute, and 'contacts'/'friends' as the
edge attribute. In~\cite{zhang2016kdd} degree
similarity is used to construct {\em prior preference matrix} $H$.

{\em Flickr-Lastfm}~\cite{zhang2016kdd-data}:
These two graphs are extracted subnetworks of the original versions~\cite{zhang2015kdd}.
The original versions contain $216K$, $136K$ users and $9M$, $1.7M$ edges respectively.
\cite{zhang2015kdd,zhang2016kdd} construct an alignment scenario for original dataset by subtracting
a small subnetwork for their ground-truth.
The two subnetworks have 12,974 nodes and 15,436 nodes,
respectively.
In extracted subnetworks, the gender of a user (male, female, unknown)
considered as the vertex attribute.
\cite{zhang2015kdd,zhang2016kdd} sort nodes by their PageRank scores to label
vertices as ``opinion leaders'', ``middle class'', and ``ordinary users''.
Edges are attributed by the level of people they connect to (e.g., leader
with leader). The user name similarity is used to construct prior preference
matrix $H$.

{\em Flickr-Myspace}~\cite{zhang2016kdd-data}:
These two graphs are extracted subnetworks of the original dataset \cite{zhang2015kdd}. Original
datasets contains $216K$, $854K$ users and $9M$, $6.5M$ edges respectively.
\cite{zhang2015kdd,zhang2016kdd} construct an alignment scenario for original dataset by subtracting
a small subnetwork for their ground-truth.
The two subnetworks have 6,714 nodes and 10,733 nodes,
respectively.
The vertex and edge attributes computed using the same process
described for Flickr-Lastfm.

{\em Facebook-Facebook}: We use Snap's~\cite{ref:snap} facebook-ego graph.
First, we randomly permute this graph and remove 20\% of the edges. Then, we
add 10\% new vertices and randomly add 10\% edges to create the second
network.

{\em DBLP (2014-2017)}: We downloaded consecutive years of DBLP graphs from 2014~\cite{ref:dnd}
to 2017~\cite{ref:dblp-site}. The ground-truth between these two graphs is created using
authors' key element. Vertices are authors and two authors have
an edge if they have co-authored information. For each publication,
DBLP records a cross-ref like `conf/iccS/2010'. We use this
cross-ref information to create vertex attributes by splitting a cross-ref by
`/' and unionizing
initial character of each word as the vertex attribute. Edge attribute between
two vertices is the mean of the publication years of co-authored
papers between two authors. The other DBLP graphs listed in Table~\ref{table:dataset},
the ones with suffixes (0) through (4), are smaller
subgraphs of the original DBLP graph, centered around highest degree
vertex.

\subsection{\sysname: Structure Assisted Partitioning}

\begin{figure}[ht]
\centering
\subfigure[Facebook]{\includegraphics[width=0.4\linewidth]{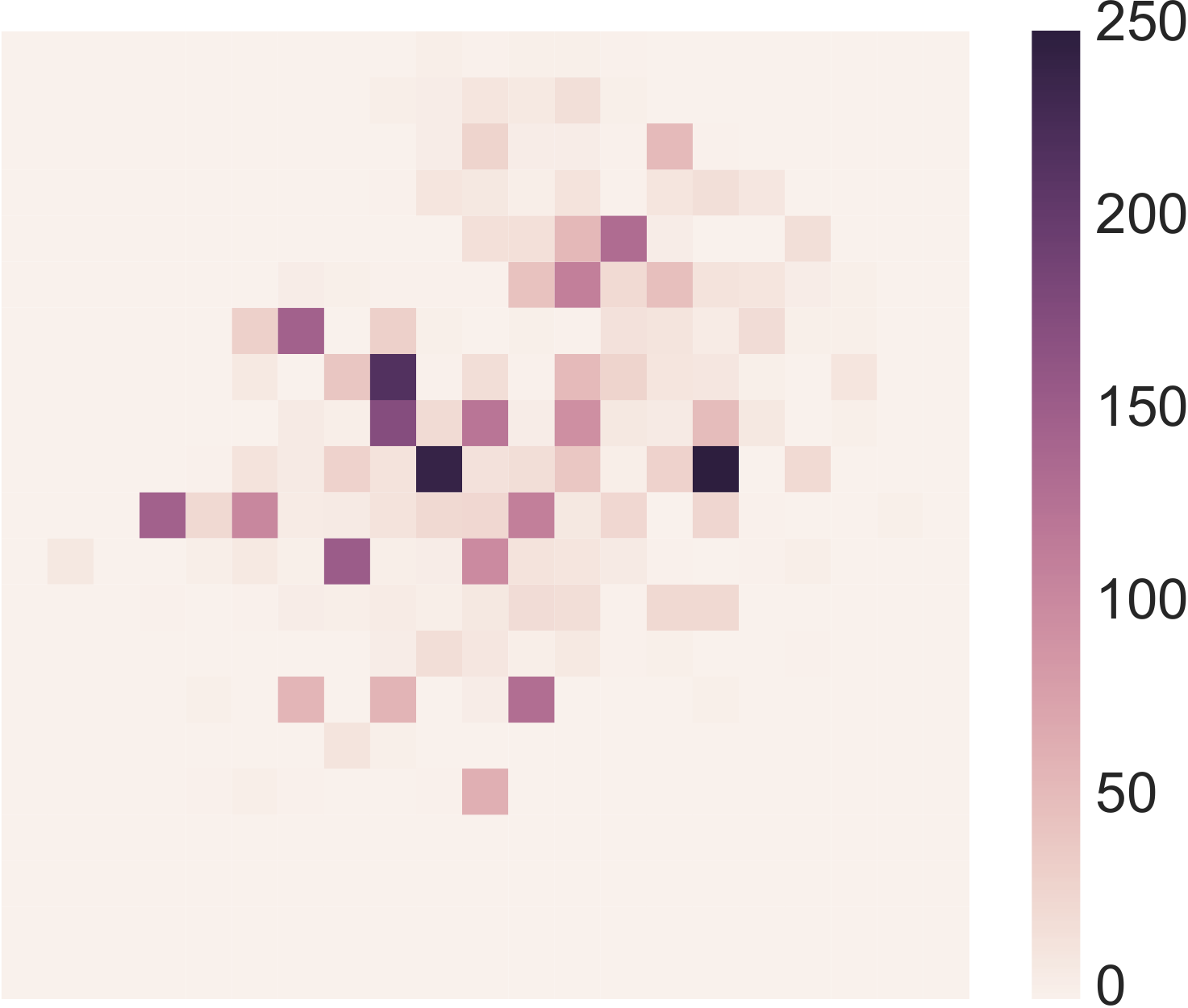}
\label{fig:hm:facebook}}
\hspace*{2em}
\subfigure[Douban]{\includegraphics[width=0.4\linewidth]{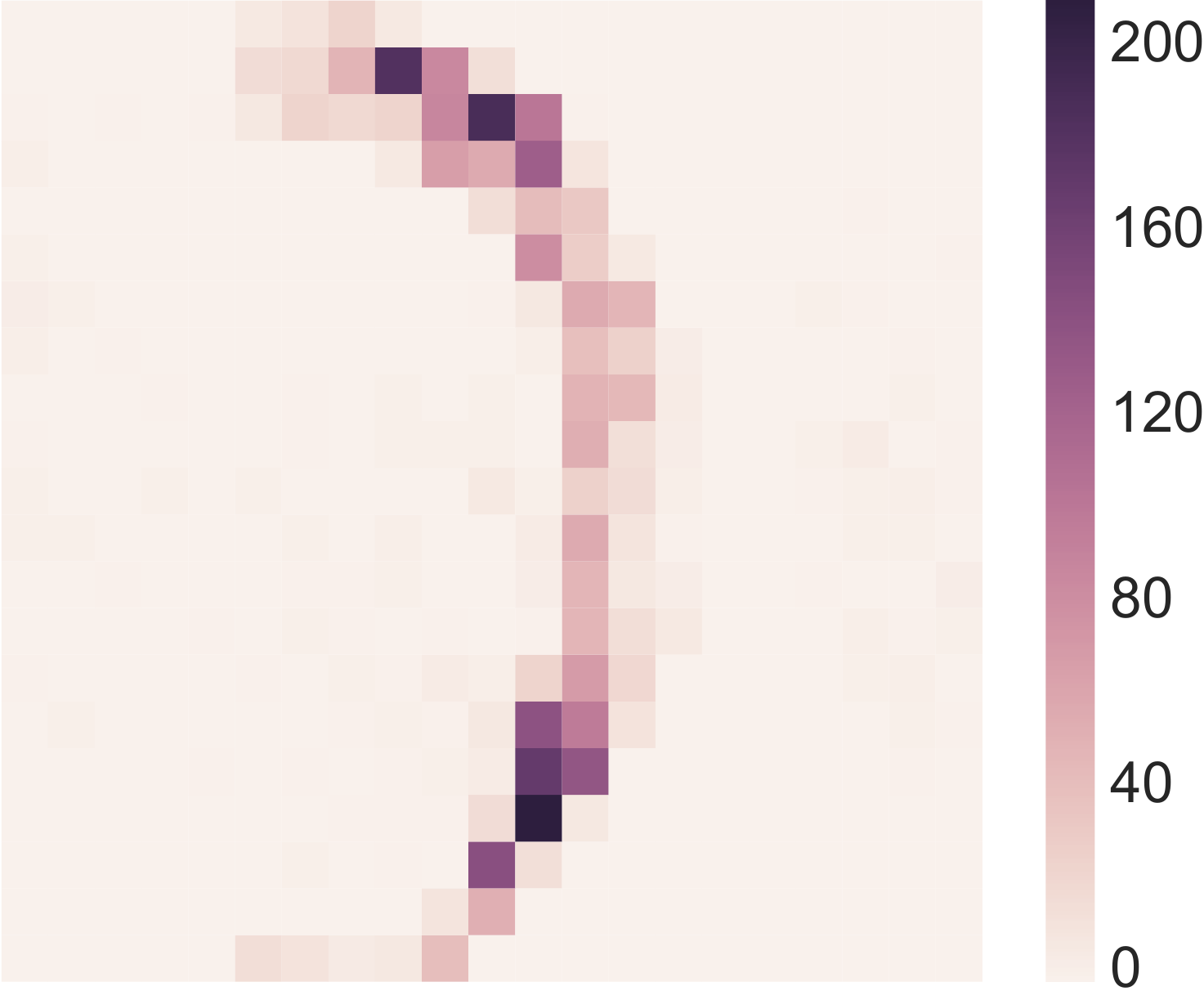}
\label{fig:hm:douban}}
\subfigure[Flickr-lastfm]{\includegraphics[width=0.4\linewidth]
{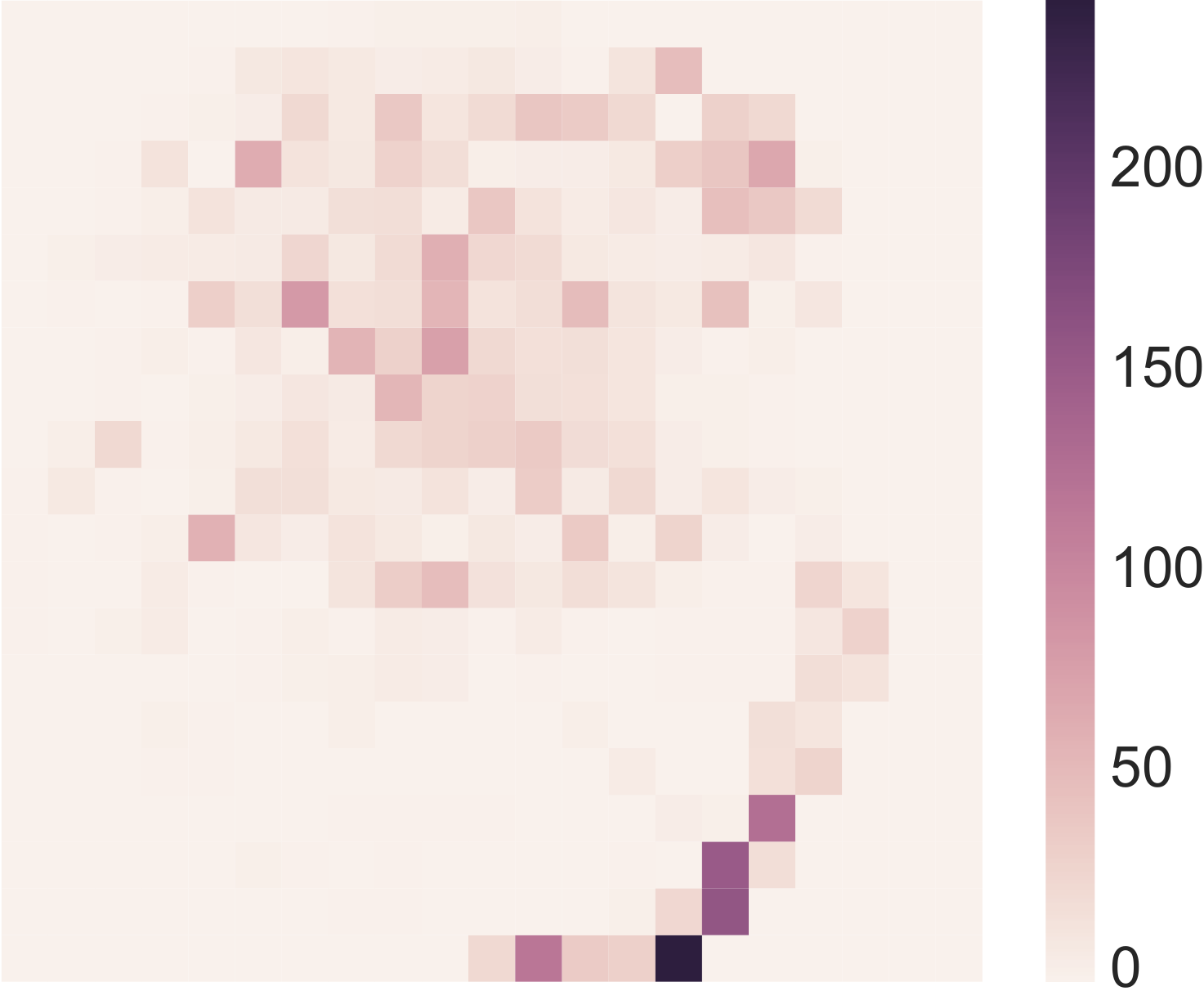}\label{fig:hm:lastfm}}
\hspace*{2em}
\subfigure[Flickr-Myspace]{\includegraphics[width=0.4\linewidth]
{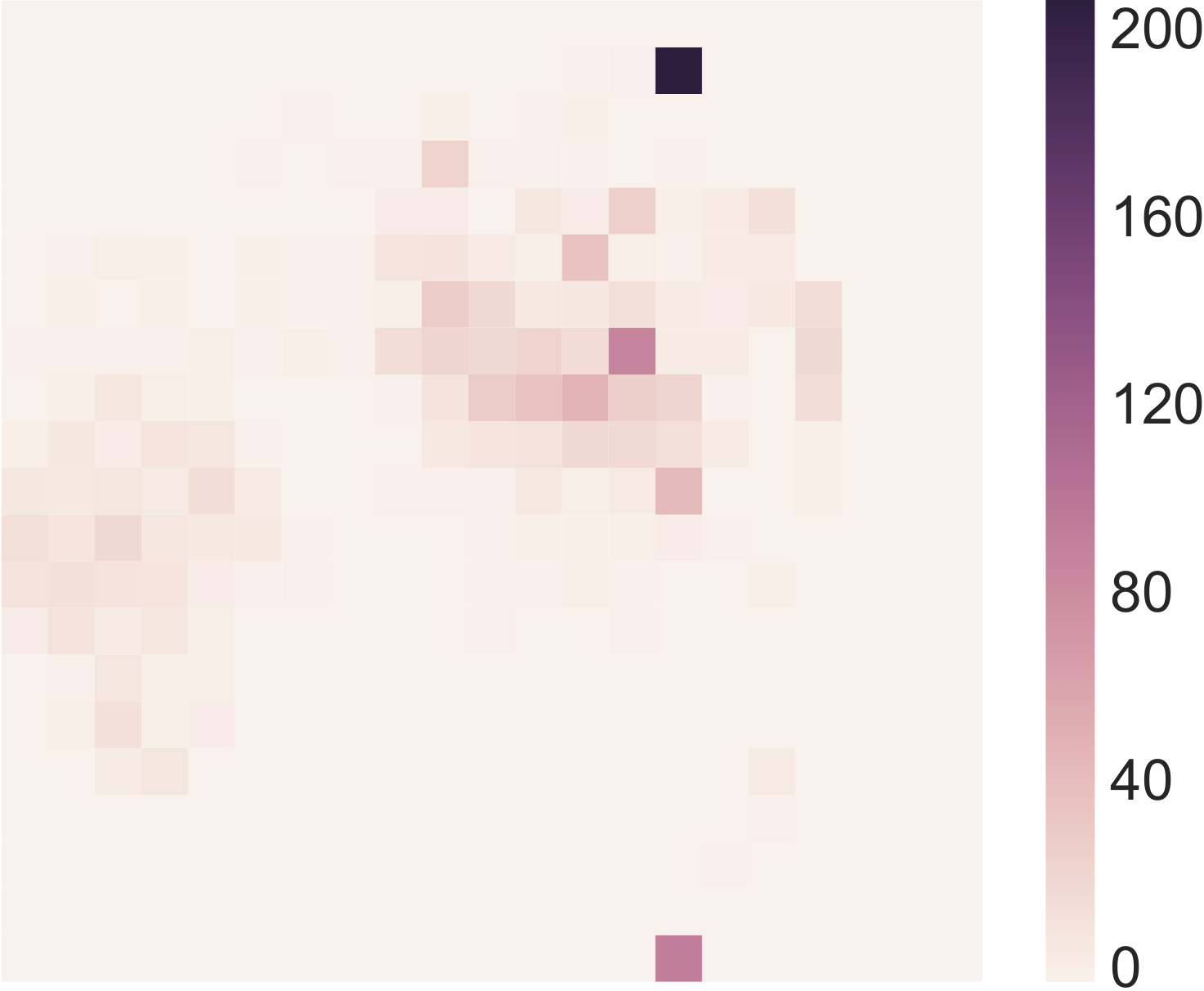}\label{fig:hm:myspace}}
\caption{Density Heat Maps}
\label{fig:hb}
\end{figure}

Figure~\ref{fig:hb} shows the density heat maps of four real-world
datasets after the vertices are positioned onto $2D$ using the techniques
presented in Section~\ref{sec:partitioning}. Each of the
sub-figures presents a square from $(-1,1)$ to $(1,-1)$.
Vertices' coordinates are found using Alg.~\ref{alg:globalpos}.
We have partitioned this space as a uniform grid with bucket sizes of
$0.1 \times 0.1$, then counted the number of vertices
in each bucket. Darker color represents higher number of vertices in that
bucket.

The first thing we observe from Figure~\ref{fig:hb} is that our
partitioning algorithm is working, that is, it enables us to partition the
vertices into different buckets by mapping them into a $2D$ and then
partitioning that plane with space partitioning techniques. As expected,
uniform density, in other words, load-balance partitioning of buckets, is
almost always impossible because of the skewness of the real-world graphs.
Therefore instead of using a grid-like partitioning, we use quadtree~
\cite{finkel1974quad} based partitioning.

\subsection{\sysname: Scope of Bucket Comparison}

In Table~\ref{tab:gain} we compare the performance of \sysname
under two settings:
first, during mapping \sysname only considers vertices in the same bucket;
second \sysname looks neighbors of each bucket for possible mappings.
In order to quantify this, we define \emph{Hit
Count} as the ratio of the number of $\mu[v] = u$ mappings considered (i.e.,
\sysname computed a similarity score between $u$ and $v$, it may or may not
map them) to the number of
such true mappings.

For the settings, we measure the {\em recall}, {\em hit count} and {\em gain}
for alignment of
{\em DBLP(2014-2016)} vs {\em DBLP(2017)} graphs. We define {\em gain} as the
ratio of the pair of vertices which we do not compute a similarity score
to the total pair of vertices.

\begin{table}[h!]
\begin{tabular}{c|c|c|c|l}
\cline{2-4}
& \multicolumn{3}{ c| }{DBLP Graphs} \\ \cline{2-4}
& $14$ vs $17$ & $15$ vs $17$ & $16$ vs $17$ \\ \cline{2-4}
&\multicolumn{3}{c|}{Without Neighbors}\\\hline
\multicolumn{1}{ |c| }{Recall} & 31\% & 32\% & 41\% \\ \cline{1-4}
\multicolumn{1}{ |c| }{Hit Count} & 40\% & 55\% & 63\% \\ \cline{1-4}
\multicolumn{1}{ |c| }{Gain} & $\approx99.97$ \% & $\approx99.98$\% &
$\approx99.98$\%\\ \cline{1-4}
&\multicolumn{3}{c|}{With Neighbors}\\\hline
\multicolumn{1}{ |c| }{Recall} & 47\% & 58\% & 66\% \\ \cline{1-4}
\multicolumn{1}{ |c| }{Hit Count} & 88\% & 92\% & 95\%   \\ \cline{1-4}
\multicolumn{1}{ |c| }{Gain} & $\approx99.85$ \% & $\approx99.85$\% &
$\approx99.86$\% \\ \cline{1-4}
\end{tabular}
\caption{Scope of bucket comparisons.}
\label{tab:gain}
\end{table}

We have following observations, first, the quality of mapping, i.e. recall,
improves with the decrease in the year differences between two graphs. This
is an expected result, for example, 2016 graph is more similar to 2017
graph than 2014 graph.
Second, the {\em hit count} rate decreases almost half
when \sysname only considers vertices within the same bucket. A similar, though
not as much, decrease is also observed in {\em recall}.
Third, {\em hit count} is sufficiently high for the second case.
Forth, the gain is very high in both cases, i.e., \sysname approximately
compares
only $1/5000$ of possible vertex pairs in the first case, and
only $1/1000$ in the second case. Based on these results, we set the
default of \sysname to consider neighbors of each bucket for possible mappings.

\subsection{\sysname: Effects of Bucket Size}

\begin{figure*}[ht]
\centering
\subfigure[Hit Count: $1^{st}$ set of Graphs]{\includegraphics[height=0.20\linewidth]{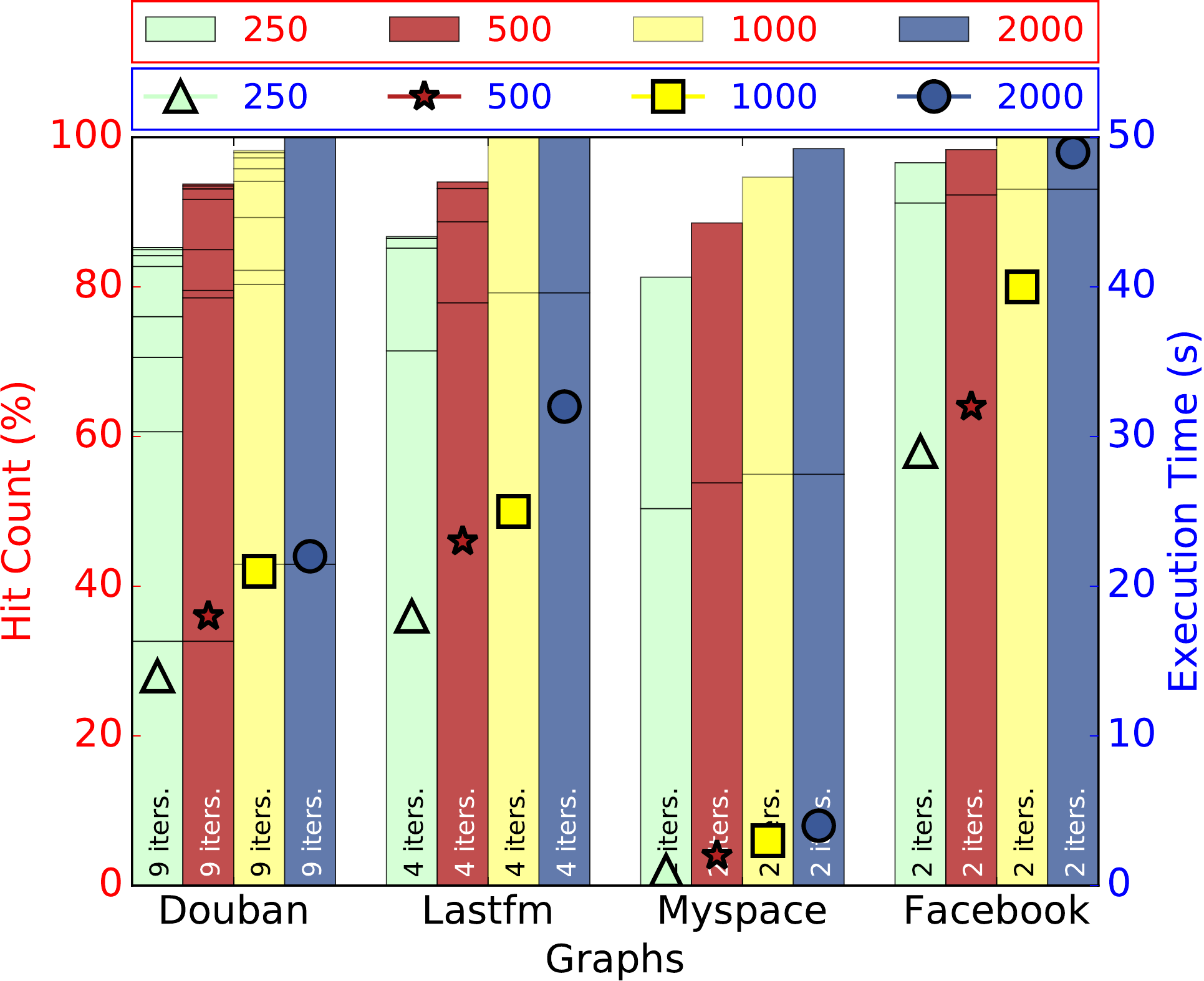}\label{fig:h1}}
\subfigure[Hit Count: $2^{nd}$ set of Graphs]{\includegraphics[height=0.20\linewidth]{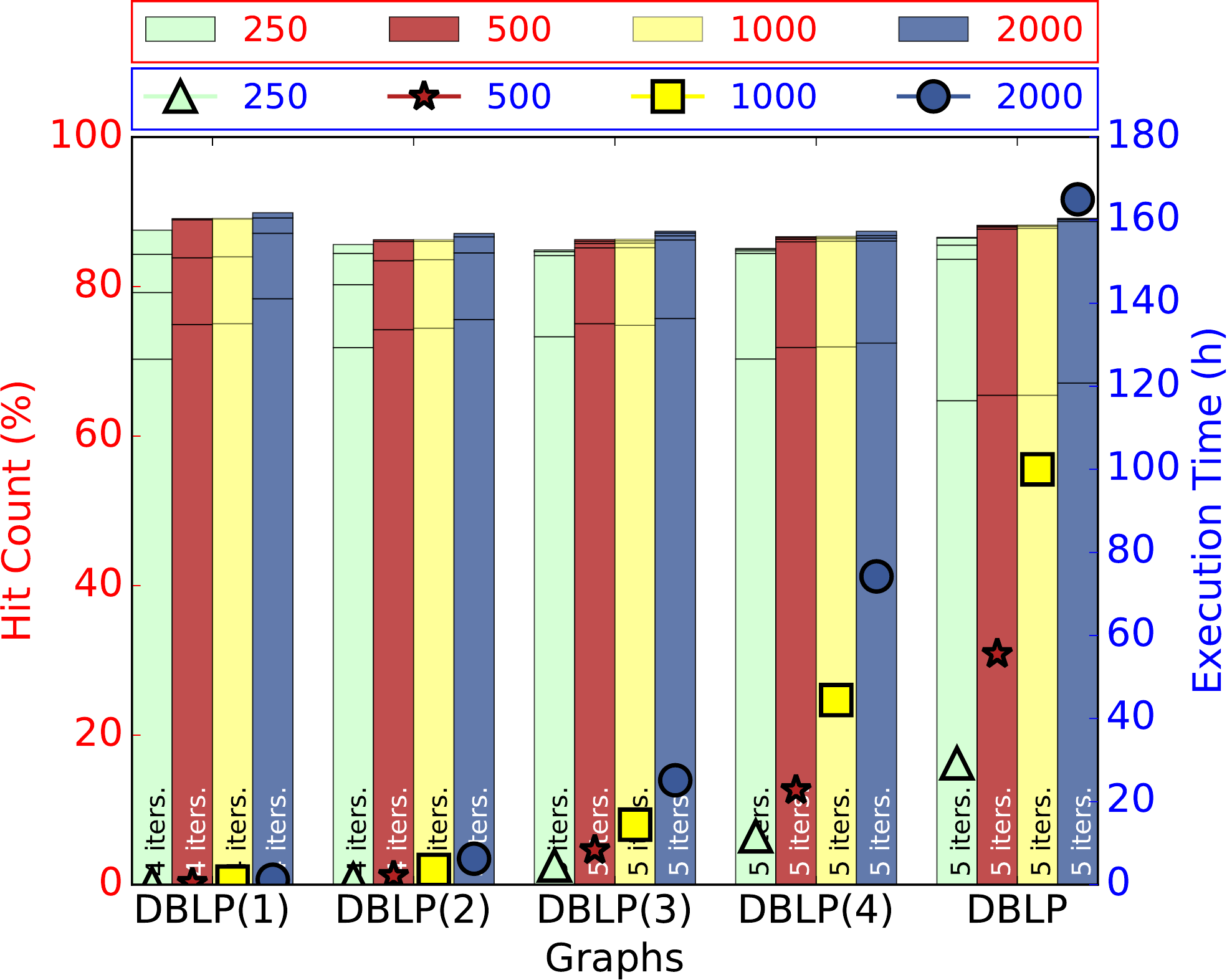}\label{fig:h2}}
\subfigure[Recall: $1^{st}$ set of Graphs]{\includegraphics[height=0.20\linewidth]{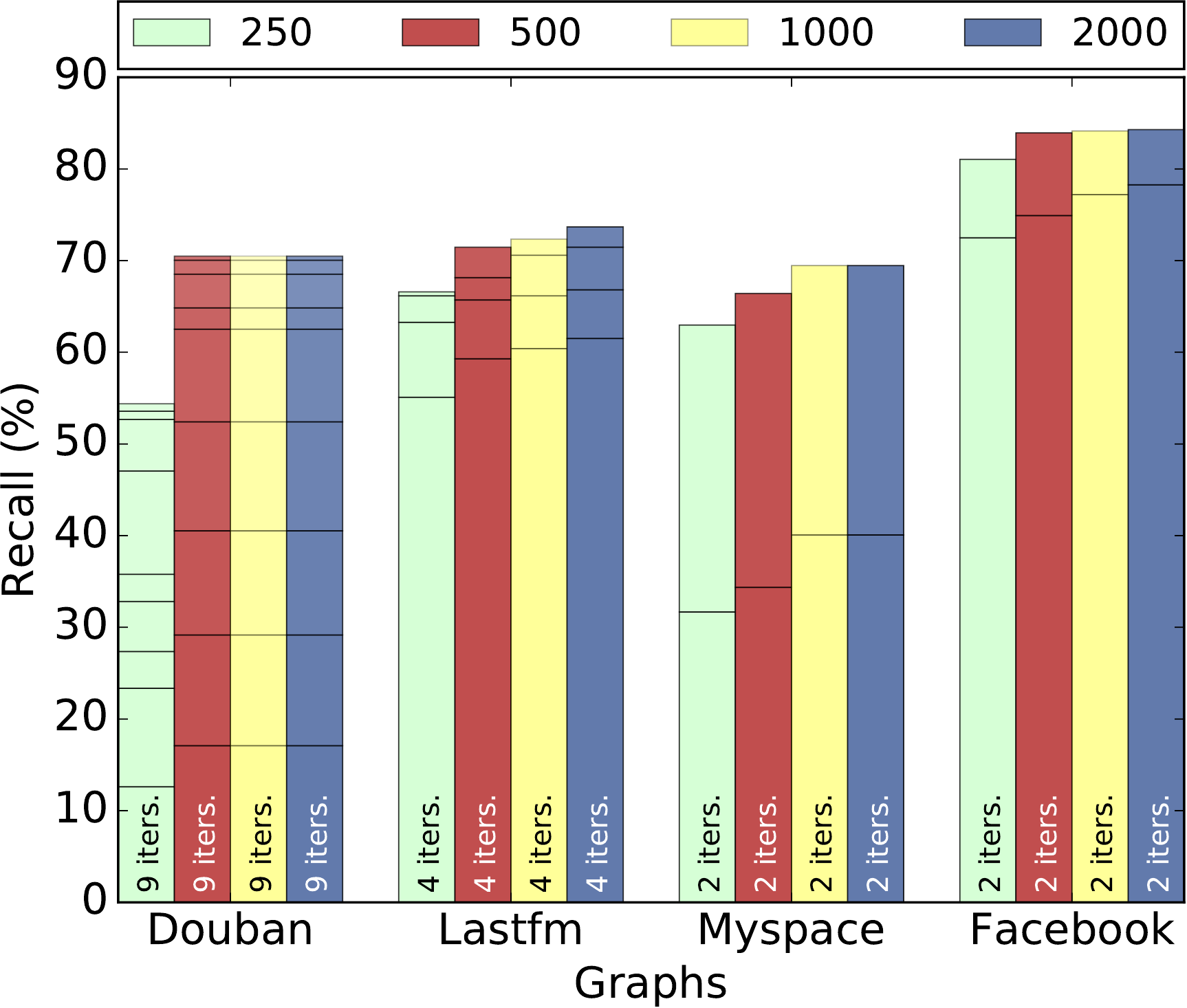}\label{fig:a1}}
\subfigure[Recall: $2^{nd}$ set of Graphs]{\includegraphics[height=0.20\linewidth]{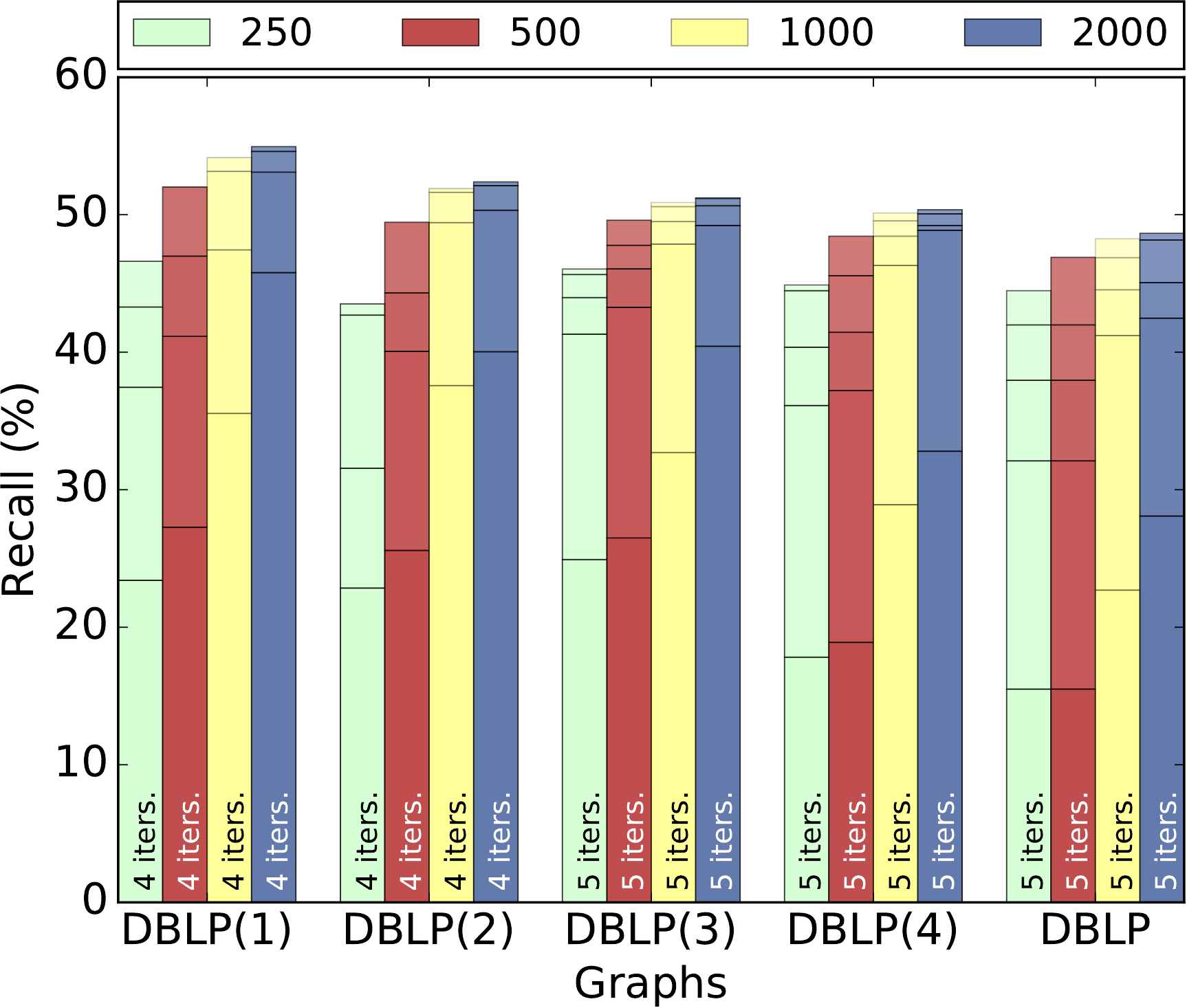}\label{fig:a2}}
\caption{Figures \ref{fig:h1} and \ref{fig:h2} plot the {\em Hit Count} (left
axis) and {\em Execution Time} (right axis), and Figures \ref{fig:a1} and
\ref{fig:a2} plot the {\em Recall}, as a function of different graph and
bucket sizes. Each bar represents a different bucket size. Number of
iterations of \sysname for each instance is printed at the bottom of each
bar, and hit count or recall at each iteration is depicted as stacked
results.}
\end{figure*}

In this section, we study
the effects of bucket size on the recall and execution time.
Figures~\ref{fig:h1} and~\ref{fig:h2} plot the {\em Hit Count} and {\em
execution} time of \sysname as a function of bucket and graph size. We observe
from Figures~\ref{fig:h1} and~\ref{fig:h2} that run time increase
is sub-linear in the size of buckets within each dataset. Quad-tree style
partitioning is one of the key factors that determine the number of comparisons
which affects the run time. When we double
the bucket size the number of buckets and average number of vertices per bucket
does not double. This explains the sub-linear trend with respect to the bucket sizes.
We observe from
Figure~\ref{fig:h2} that number of edges affects runtime because it affects
the complexity of our similarity function. For instance, running time 
increases in average about
$5 \times$ between DBLP(0) and DBLP(1) graphs and $2.3 \times$
between DBLP(4) DBLP, while the number of edges increase in average about 
$1.9 \times$ and $1.4 \times$ respectively.
We also
observe from Figures~\ref{fig:h1} and~\ref{fig:h2} that {\em Hit Count}
slightly increases with increasing bucket sizes.

Briefly,
recall is the ratio of the correct mapping found by \sysname to the number
of ground truth mapping. Figures~\ref{fig:a1} and~
\ref{fig:a2} show the trend in recall as a function of different bucket sizes
for different graphs.
From the figures, we observe that increasing the
bucket size increases the recall, but there is a diminishing return as
expected. Recall increases about $8\%$ on the average with increasing bucket
sizes from $250$ to $2000$ and only $4\%$ when
bucket sizes from $500$ to $2000$. Based on these results we picked bucket
size $500$ as our default for further experiments.

\subsection{Comparison against state-of-the-art}
Here, we compare our proposed algorithm, \sysname, with four
state-of-the-art mapping algorithms:  {\tt IsoRank}~\cite{singh2007pairwise},
{\tt Klau}~\cite{klau2009bmc}, {\tt NetAlign}~\cite{bayati2009icdm},
and {\tt Final}~\cite{zhang2016kdd}.

\begin{figure*}[ht]
\centering
\subfigure[Anchors are not known]{\includegraphics[width=0.32\linewidth]
{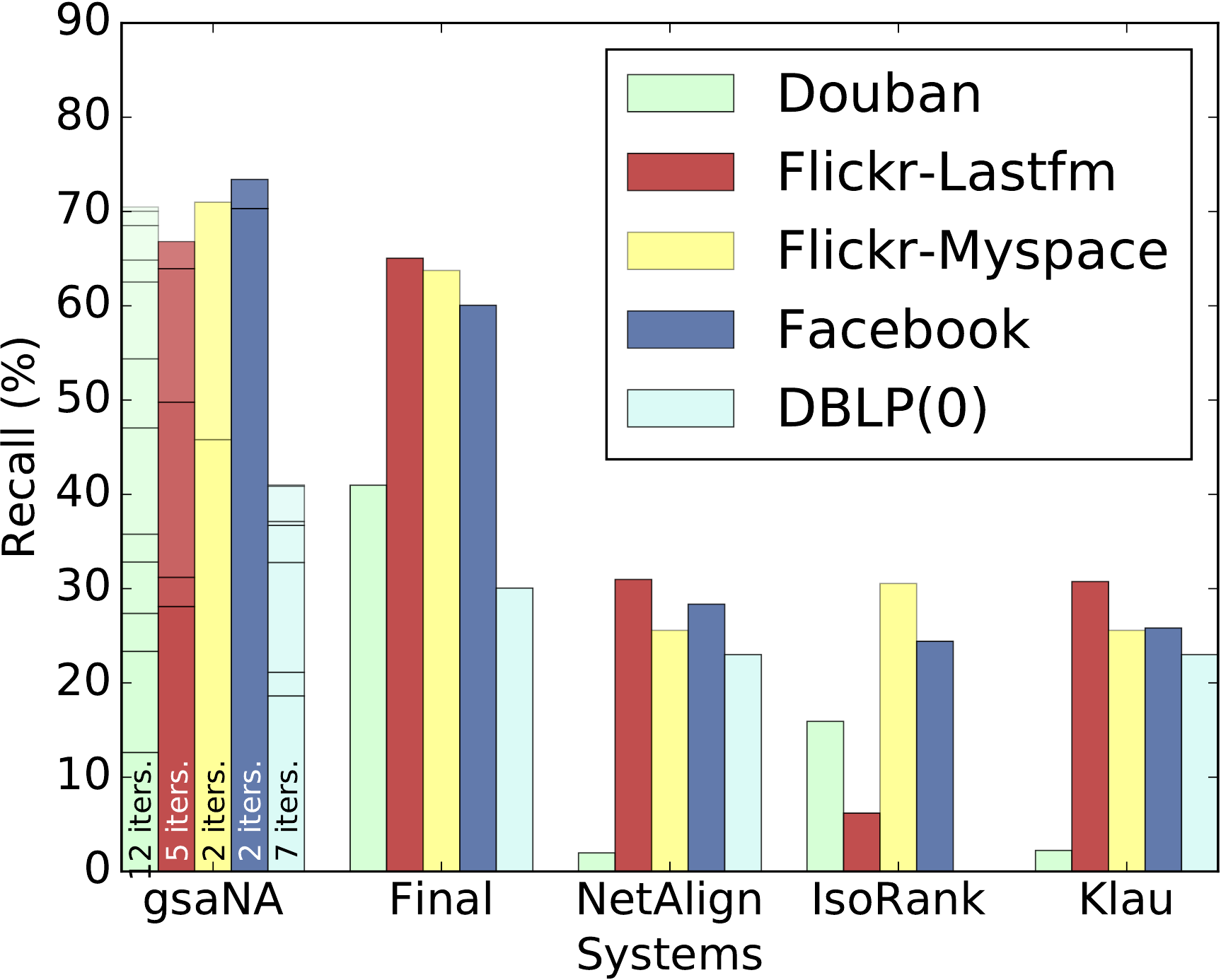}\label{fig:attr}}
\hspace*{.5em}
\subfigure[Anchors are known]{\includegraphics
[width=0.32\linewidth]{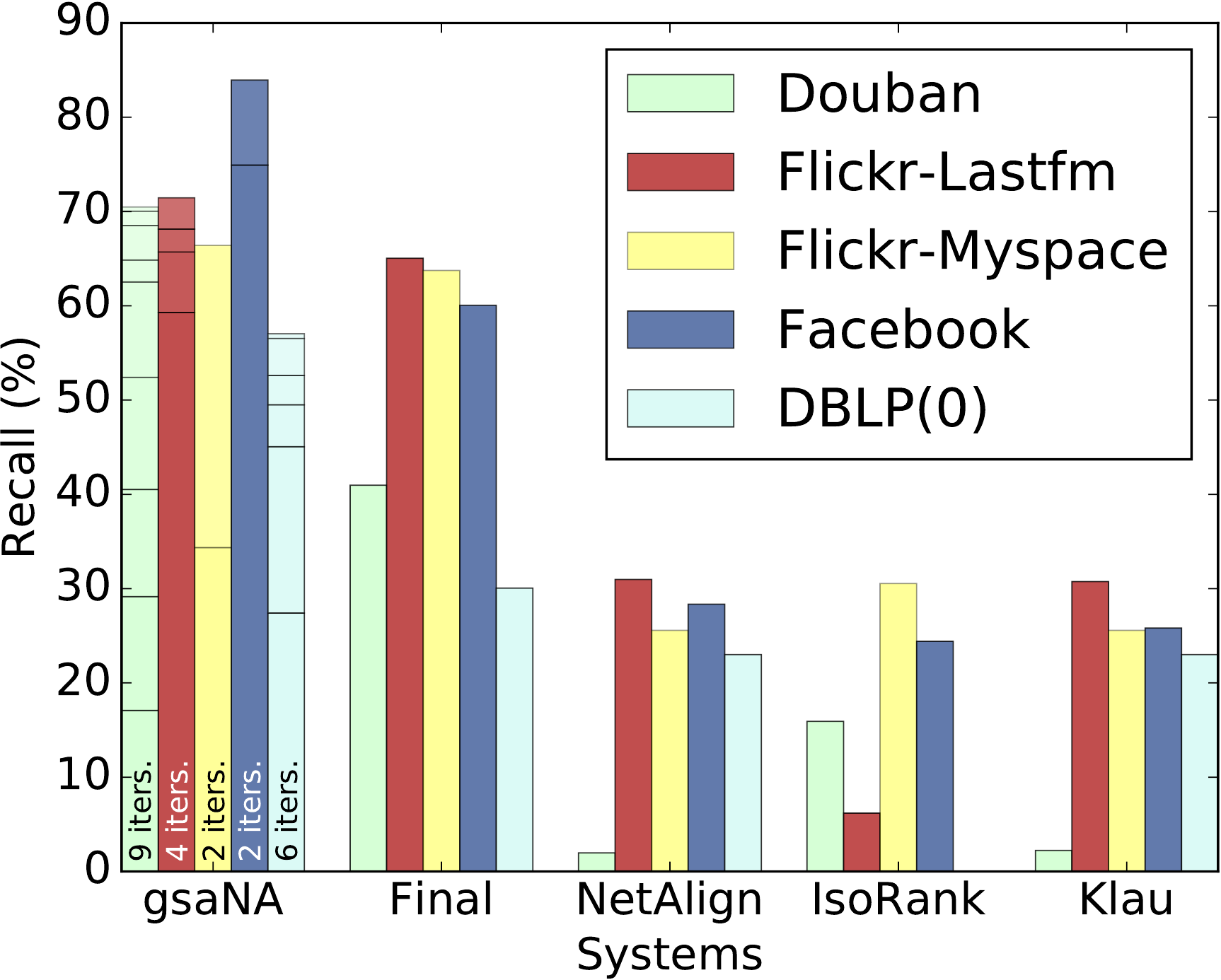}\label{fig:akn}}
\hspace*{.5em}
\subfigure[Execution Time]{\includegraphics[width=0.32\linewidth]{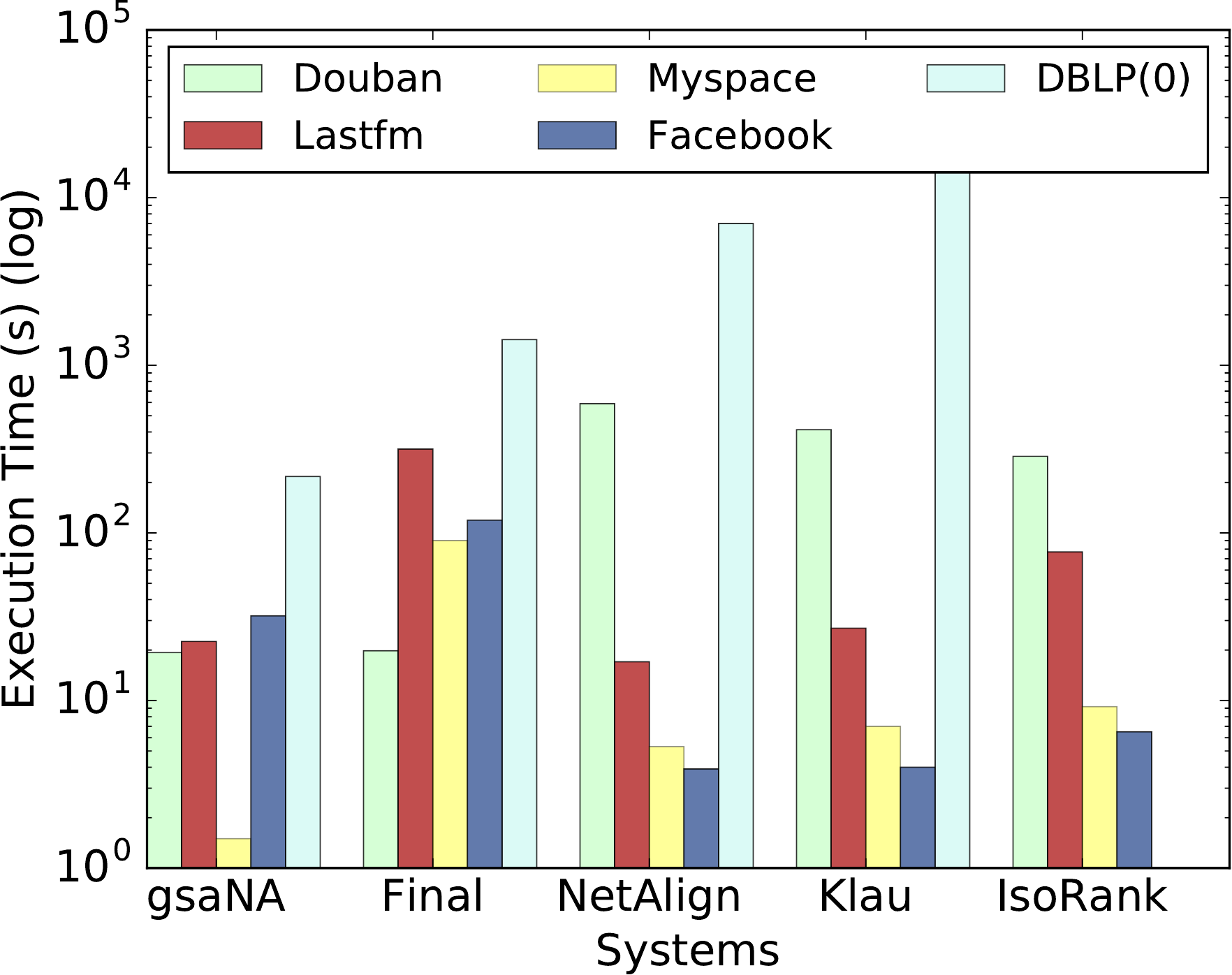}\label{fig:exe}}
\caption{Figures \ref{fig:attr} to \ref{fig:akn} plot the {\em Recall}
of the systems under conveyed scenarios.
Figure \ref{fig:exe} plots the {\em Execution Time}.
In each plot (left axis)
represents {\em Recall} or {\em Execution Time }. Each bar represents a different
graph.
}
\end{figure*}

\begin{figure}[ht]
\centering
\subfigure[Structural Noise]{\includegraphics[width=0.45\linewidth]{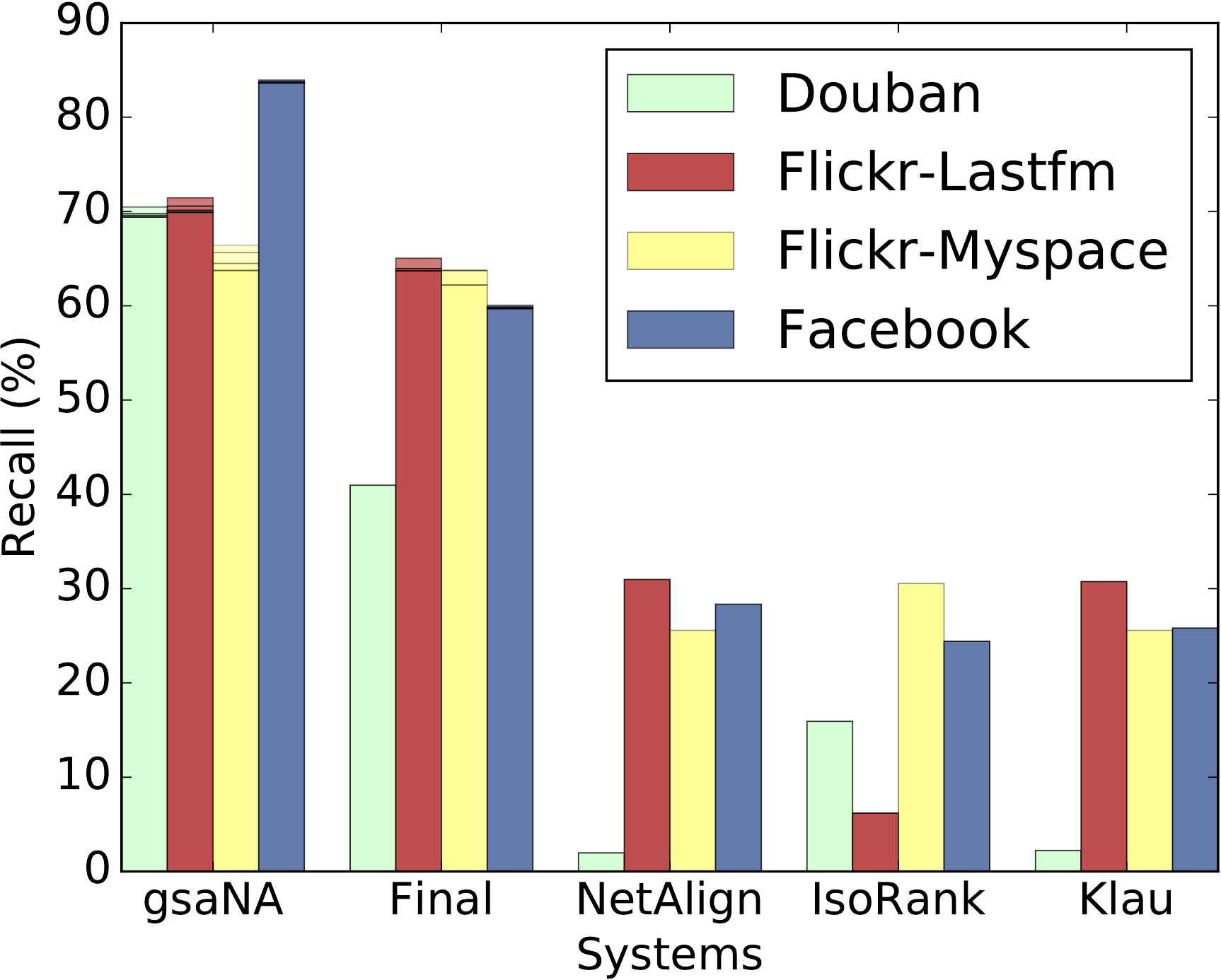}\label{fig:str-err}}
\hspace*{1em}
\subfigure[Attributed Noise]{\includegraphics[width=0.45\linewidth]{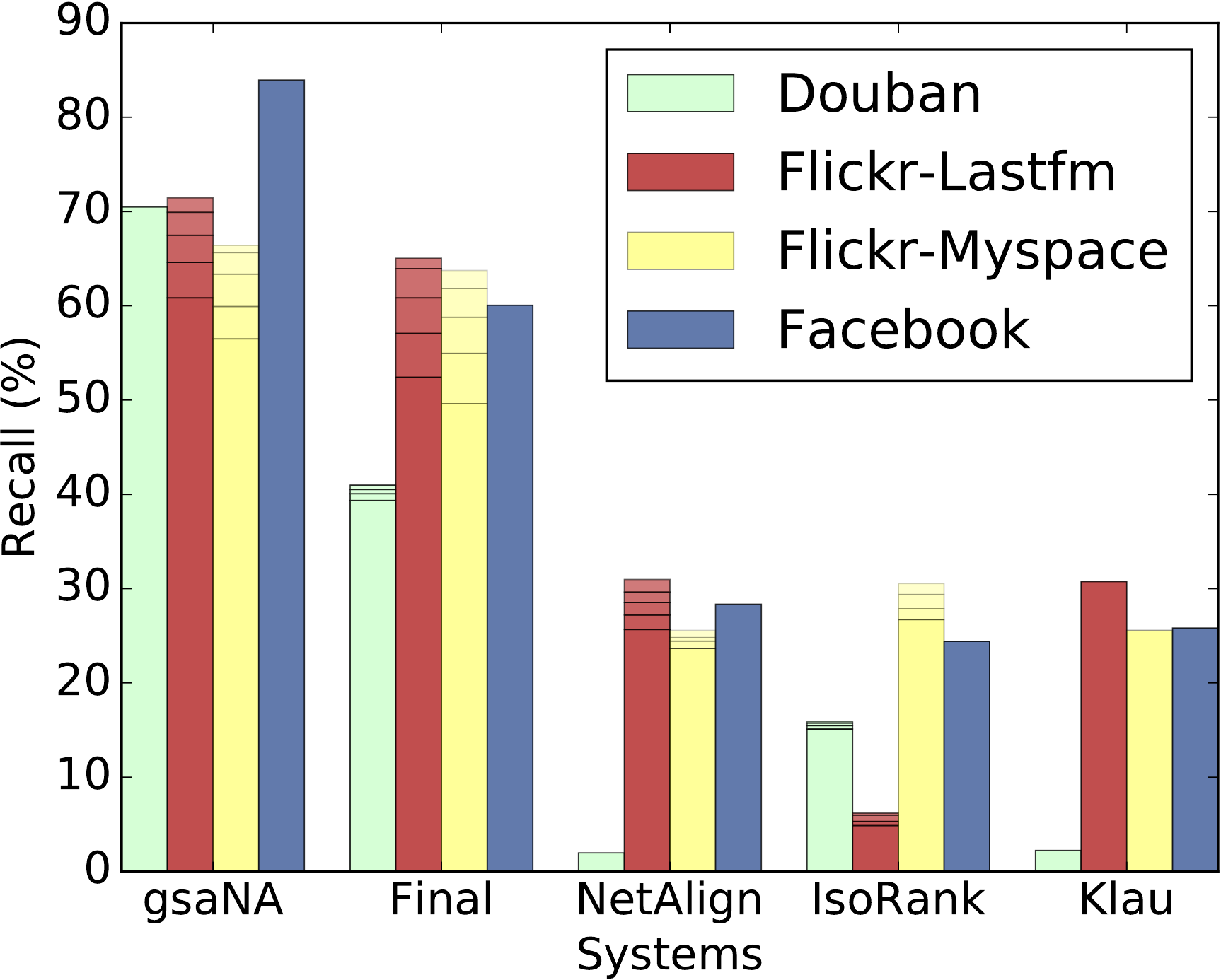}\label{fig:attr-err}}
\caption{ {\em Recall} under
noise. In each plot (left axis) represents {\em Recall} or {\em Execution Time }. Each
bar represents a different graph.
}
\end{figure}

In the experiments presented in Section~\ref{subsec:attr} to~\ref{subsec:noise}
(Figures~\ref{fig:attr}-\ref{fig:attr-err} respectively) we also take
additional metadata information, such as vertex and edge attributes, types,
etc., whenever it exist.
{\tt NetAlign}~\cite{bayati2005isit} and {\tt Klau}~\cite{klau2009bmc}
require an additional bipartite graph, representing the similarity scores
between two input graphs' vertices.
{\tt Final}~\cite{zhang2016kdd}'s goal is to leverage the additional
metadata information
and improve {\tt IsoRank}~\cite{singh2007pairwise}. Therefore, in these experiments
{\tt Final}'s~\cite{zhang2016kdd} {\em prior preference matrix H} is used for
Douban, Flickr-Lastfm and Flickr-Myspace graphs for all other algorithms. We
have used $H$ as \sysname's $C_V$ for Flickr-Lastfm and Flickr-Myspace
graphs, since they reflected vertex attribute similarity, and vertex
attributes were not provided separately. In Facebook, each
vertex considered as possible mapping between top similar (computed as
$\sigma = \tau \times \Delta$, see Section~\ref{sec:similarity}) $s$
vertices, where
$s$ is randomly selected number in the range of $[5,15]$.

In DBLP(0),
first a similarity matrix
is generated using $C_V$ and then all elements smaller than 0.9
set as 0. Both for Facebook and DBLP(0) after deciding possible
mappings we have also added ground truth for not to miss any information.
In order to be fair, and help to improve {\tt IsoRank}'s result, we also set
its similarity matrix's elements corresponding to 0 elements in H as 0
as well.

\subsubsection{Anchors are not known}
\label{subsec:attr}

Figure~\ref{fig:attr} plots the results where
we assume anchors are not given to \sysname by the
user, and \sysname computes anchors as described
in Sec.~\ref{subsec:anchorselection}.
As seen in the figure, \sysname outperforms all of the algorithms in terms of
recall.
On the average,
\sysname produces about $1.3\times$
better recall than {\tt Final}, $9\times$  better recall
than {\tt NetAlign}, $5\times$ better recall than {\tt IsoRank} and
$8\times$ better recall than {\tt Klau}. However, {\tt NetAlign}
and {\tt Klau} performs really poor on Douban dataset, therefore if we omit
this dataset \sysname produces $2.3\times$ and $2.4\times$
better results than {\tt NetAlign} and {\tt Klau}, respectively.

\subsubsection{Anchors are known}
\label{subsec:anch}

Figure~\ref{fig:akn} presents the results where the {\em anchor set} is given
by the user.
We set these anchors' similarity score as 1.0 in all the other
algorithms we compare too.
We observe that \sysname's recall increases in Facebook and DBLP(0)
graphs because wrong initial anchor mapping
are corrected. However, Flickr-Myspace graphs' recall slightly decreases
\sysname produces about $1.4\times$
better recall than {\tt Final}, $9\times$  better recall
than {\tt NetAlign}, $5\times$ better recall than {\tt IsoRank} and
$\approx 8\times$ better recall than {\tt Klau}. Same as previous experiment
if we omit Douban dataset \sysname produces $2.6\times$ and
$2.7\times$ better results than {\tt NetAlign} and {\tt Klau},
respectively.

\subsubsection{Execution Time}
\label{subsec:exec}

Figure~\ref{fig:exe} displays the execution time results, in log-scale, of
the algorithms we compare.
In this figure, the pre-processing time of computing
similarity bipartite graph for {\tt NetAlign} and {\tt Klau} and $H$
matrix for {\tt Final} is not included (we have directly used the $H$ matrix
provided with {\tt Final} implementation). We would like to note
that, computation of those similarity scores requires a significant time.
Another point we need to remind, \sysname is written in C++ while the other
algorithms are implemented in Matlab. Hence, it may not be appropriate to
compare individual absolute results, but still these results should be
good to provide some insights to trends of the execution time.

For small graphs, all algorithms are ``fast enough'' to use in practice.
However, for DBLP(0), which the smallest of our DBLP graphs, as you can see
other algorithms becomes orders of magnitudes slower. \sysname can solve our
largest DBLP graph, 32 times larger than DBLP(0), almost with same time they
take for DBLP(0).

\subsubsection{Effect of Errors}
\label{subsec:noise}

In Figure~\ref{fig:str-err} we present results when there is structural error
in the input graphs. We randomly remove 5\%, 10\%, 15\% and 20\% of the
edges from both graphs, then for each case we re-run the systems. Since we
observed only small amount of change in the results, and recall of mapping
decreased with increasing error rate, in all experiments, we simply plotted
them as a stacked bar results, that is there are 4 horizontal lines in each
bar depicting 5\%, 10\%, 15\% and 20\% error, from top to bottom.
We expect, at some point, \sysname will be effected from structural
error because eventually shortest paths are going to change, and
hence partitioning. However, as seen in the figure, removing edges
up to 20\% did not significantly change the partitioning because the
results doesn't significantly changed, i.e. still \sysname has good
hit count ratio.

In Figure~\ref{fig:attr-err} we present results when there are errors in
attributes. Since we had used $H$ matrices provided by {\tt Final} as
our attribute similarity, basically we randomly changed 5\%, 10\%,
15\% and 20\% of the non-zero elements of $H$. And for each
case we re-run all the algorithms, except {\tt  Klau}~\cite{klau2009bmc},
since the errors in attributes do not affect it. As expected other systems'
recalls, including that of \sysname, decrease when we increase
the noise. We also observe that interestingly while removing edges randomly
doesn't affect IsoRank, adding noise to its similarity matrix changes its
final recall. This experiment, as expected showed that largest changes in the
recall were in \sysname and {\tt Final}, especially in Flickr-Lastfm and
Flickr-Myspace data sets, since those are the algorithms that incorporates
the attribute similarity.

\section{Conclusion}
\label{sec:conclusion}

We have developed an iterative graph alignment framework called \sysname,
which leverages the global structure-based vertex positioning technique to
reduce the problem size, and produces high quality alignments that
outperforms the state-of-the-art. As the graph sizes increases, the runtime
performance of the proposed algorithm becomes more pronounced, and becomes
order of magnitudes faster than the existing algorithms, without a
significant decrease in the performance.
As a future work, our goal is parallelize \sysname to take advantage of
multi-node and/or multi-core architectures. Many parts of the algorithm, like
initial distance computations from multiple anchors, and pairwise similarity
computation, which are the most two time consuming part of the \sysname, can
be easily parallelized. We also would like to explore techniques to extend
\sysname to solve multi graph alignment problem.

\section*{Acknowledgment}
We would like to extend our gratitude to Dr. Bora U\c{c}ar for his valuable
comments and feedbacks for the initial draft of this manuscript.

\clearpage

\bibliographystyle{ACM-Reference-Format}
\bibliography{paper-arxiv}

\end{document}